\documentclass[aps,pra,onecolumn,10pt,notitlepage,superscriptaddress,showkeys,nofootinbib,amsmath,amssymb]{revtex4-2}

\usepackage[utf8]{inputenc}
\usepackage{graphicx}

\usepackage{url}
\usepackage[hidelinks]{hyperref}
\DeclareMathOperator{\sinc}{sinc}
\DeclareMathOperator{\Var}{Var}
\newcommand{\G}{\mathfrak g}

\begin{document}
\title{Quantum Startup, Recoil Statistics and Collective Gain in Free-Electron Lasers}
\author{G. Dattoli}
\thanks{Corresponding author}
\affiliation{ENEA, Nuclear Department, Frascati Research Center, Via E. Fermi 45, 00044 Frascati (Rome), Italy}
\author{A. Curcio}
\affiliation{SBAI Department, Sapienza University, Via Scarpa 14, 00161 Rome, Italy}
\author{E. Di Palma}
\affiliation{ENEA, Nuclear Department, Frascati Research Center, Via E. Fermi 45, 00044 Frascati (Rome), Italy}
\begin{abstract}
We examine the conditions under which quantum signatures can be distinguished from classical dynamics and
statistical fluctuations in a free-electron laser. Starting from minimal coupling and the
Bambini--Renieri frame, we derive the momentum--photon exchange Hamiltonian and its Raman--Nath ladder.
An interaction-picture expansion retains detuning exactly and organizes recoil corrections through the
mismatch between neighboring orders. Its validity depends on the accumulated recoil phase over the
populated ladder, rather than on elementary recoil alone. Numerical checks and Gaussian entrance packets
illustrate the effects of preparation and resonance averaging. We distinguish number-state exchange
amplitudes from predictions for a coherent optical seed, and joint electron--field coherence from reduced
radiation coherence. The Wigner representation connects finite recoil translations with classical
transport and identifies the information required for bunching. A cavity channel then provides the basis
for comparison with the recoil-resolved oscillator proposed by Kling and Giese, including the
optical-undulator preparation, stimulated back-action and narrow resonance acceptance. Space charge,
spontaneous diffusion and FEL-induced energy spread are compared with recoil and high-gain bandwidth
scales. The coherent undulator pump is essential to exclude artificial zero-recoil antibunching from a
number-state preparation. Startup and excess photon variance alone do not establish a quantum signature:
the specified observable must remain distinguishable from a classical stochastic baseline under
compatible preparation, propagation and detection conditions.
\end{abstract}
\maketitle

\section{Introduction}

In this article we discuss a recurring question in free-electron laser (FEL) theory: under what conditions can
quantum signatures of the interaction be observed? The question dates back to the foundations of the subject.
Madey described stimulated emission in a periodic magnetic field in terms of the exchange of radiation photons
and undulator pseudophotons \cite{ref1}. This language connects the microscopic interaction to stimulated
gain, but does not by itself identify an experimentally resolvable departure from classical FEL dynamics.

Independent early quantum descriptions include the theory of Becker and Mitter \cite{ref2} and the
strong-field stimulated-process analysis of McIver and Fedorov \cite{ref3}. The literature has developed
considerably, from the early quantum analyses and the previous references collected therein \cite{ref4,ref5},
to studies of coherent states, classical correspondence and quantum interaction \cite{ref6,ref7,ref8}. Becker
and McIver developed a fully quantized many-particle treatment of small-signal gain and photon statistics
\cite{ref9}; their subsequent review extends the discussion to a class of free-electron devices, including
optical wigglers \cite{ref10}. Gea-Banacloche developed a many-particle quantum treatment connecting large
gain, saturation and photon-number moments \cite{ref11}; the review by Friedman, Gover, Kurizki, Ruschin and
Yariv places spontaneous and stimulated emission in the broader framework of quasifree-electron radiation
\cite{ref12}. Early work on photon statistics and squeezed radiation also illustrates the diversity of
quantum-optical questions posed for the FEL \cite{ref13,ref14}. These early analyses already connect recoil,
spontaneous startup, stimulated amplification and the statistical preparation of the beam and field. Later
work addresses high-gain quantum corrections \cite{ref15} and collective amplification when discrete recoil
becomes dynamically important \cite{ref16}. Kling and collaborators identified a recoil-selective two-state
regime and its quantum-optical interpretation \cite{ref17}, then developed a momentum-space Dicke description
of exponential gain \cite{ref18}. Their subsequent long-time analysis distinguishes photon-number means from
preparation-dependent fluctuations and examines experimental requirements \cite{ref19}. These contributions
concern related, but distinct, levels of description. Relevant references will be quoted in the course of the
discussion. The Bambini--Renieri (BR) moving-frame formulation supplies a particularly useful bridge between
the relativistic laboratory description and the effective longitudinal interaction \cite{ref20,ref21}.

In the following, we make extensive use of the quantum analysis of the FEL developed within the
Bambini-Renieri Hamiltonian framework. Early treatments reproduced the classical small-signal gain and
clarified the difficulty of isolating quantum signatures. Subsequent contributions established a broad
theoretical basis for a dedicated experimental programme. Raman-Nath ladder equations and recoil expansions
appeared in several early treatments and became widely used; the present discussion does not assign these
tools exclusively to a single contribution. The references cited here trace the development from low-gain
quantum dynamics to collective high-gain models and identify the present experimental bottlenecks
\cite{ref2,ref3,ref4,ref5,ref6,ref7,ref8,ref9,ref10,ref11,ref12,ref13,ref14,ref15,ref16,ref17,ref18,ref19}.

We distinguish single-particle quantum effects from collective quantum effects. At the single-electron level,
quantized momentum exchange, the energy differences between recoil orders, and electron--photon entanglement
can change transition probabilities and conditional radiation statistics. These effects can be studied before
introducing beam amplification. Collective quantum effects arise when such microscopic transitions participate
in the self-consistent evolution of many electrons coupled to a common radiation field: recoil can then change
the gain spectrum, accessible momentum populations and correlations of the amplified field. Multiplying a
single-electron emission probability by the electron number does not establish collective quantum dynamics.
Conversely, bunching and collective gain already exist in classical FEL theory and are not, on their own,
evidence of quantum recoil.

The observability of a quantum correction must be assessed against competing effects: electron energy spread
and emittance, undulator imperfections, finite-electron shot noise, and stochastic energy changes caused by
spontaneous radiation in the undulator. The importance of each effect depends on the operating regime. A
quantum description of emission is necessary at the microscopic level, but observing radiation or startup does
not alone demonstrate a nonclassical signature. Amplifier and multimode analyses illustrate why the optical
state and its statistics must be considered together with the beam \cite{ref22,ref23}. The appropriate
comparison is therefore between quantum and classical predictions for the same measured quantity, with the
same beam preparation, interaction and detection conditions.

The plan of the article follows this distinction. We first establish the Hamiltonian, its reference frame and
its operator structure, explaining how the prescribed undulator enters the reduced interaction. The
interaction-picture treatment then derives detuning and recoil, and identifies the prescribed-current solution
as a useful reference. We develop the Raman--Nath momentum--photon ladder and its perturbative recoil
corrections before discussing modulation, bunching and beam averaging. This order separates elementary quantum
transitions from the collective response that may amplify or obscure them. Only afterwards do we introduce the
resonator and compare its dynamics with the recoil-resolved, approximately two-state oscillator of Kling and
Giese \cite{ref24}. The comparison asks which quantum signatures survive fresh-electron pumping and cavity
loss, recognizing the innovative experimental perspective of the oscillator proposal. The final appendices
compare low-gain quantum dynamics with high-gain SASE and establish the shot-noise baseline for interpreting
startup and radiation statistics.

The contribution of the present work is a unified, preparation-sensitive account of recoil observability. We
formulate the detuning-exact local-mismatch hierarchy, validate its second-order amplitudes against direct
ladder propagation, and connect it to reduced optical coherence and beam averaging. We then use these
distinctions to assess resonance acceptance and photon-statistical diagnostics in an optical-undulator
oscillator. These results delimit the inference available from specified observables; they do not constitute a
new collective high-gain theory or a demonstration of experimental feasibility.

\section{Relativistic starting point, gauge choice and FEL frame}

The laboratory-frame starting point is the spin-independent, positive-energy minimal-coupling Hamiltonian for
an electron of charge $-e$,

\begin{equation}
\hat H=\sqrt{m^2c^4+c^2[\hat{\boldsymbol p}+e\boldsymbol A_u(\hat{\boldsymbol r})+e\hat{\boldsymbol A}_r(\hat{\boldsymbol r})]^2}+\hat H_{\rm rad}.
\label{eq:1}
\end{equation}

Here $m=m_{e}$ is the invariant electron rest mass; $c$ the speed of light, $\boldsymbol{A}_{u}$ the
prescribed undulator vector potential and $\hat{\boldsymbol{A}}_{r}$ the quantized radiation vector potential.
Spin and pair production are outside this effective description. We use the Coulomb (radiation) gauge,
$\nabla\cdot\boldsymbol{A}=0$, with $\Phi=0$ for the transverse laboratory problem. Physical transverse photon
degrees of freedom are explicit in this gauge; gauge-invariant observables do not depend on the choice. A
subsequent boost and reduction must transform the fields consistently rather than retain laboratory potentials
unchanged.

The reduction to one longitudinal degree of freedom must be stated carefully. In the laboratory the electron
has a large longitudinal momentum, but the undulator generates an oscillating transverse mechanical momentum.
For a planar undulator its characteristic magnitude is

\begin{equation}
\frac{p_\perp}{mc}\sim K,\qquad K=\frac{eB_u\lambda_u}{2\pi mc}.
\label{eq:2}
\end{equation}

Here $B_{u}$ and $\lambda_{u}$ are the laboratory peak field and period. Polarization and averaging introduce
the familiar convention-dependent factors. The electron--radiation coupling remains possible when the reduced
dynamics is written only in terms of $z$ and $p_{z}$: the transverse motion has first been generated by the
undulator and eliminated in favor of an effective longitudinal interaction.

This resolves an apparent difficulty in Eq. (1). If $\hat{\boldsymbol{p}}=\hat{p}_{z}\boldsymbol{e}_{z}$ and
both vector potentials are transverse, $\hat{\boldsymbol{p}}\cdot\boldsymbol{A}=0$. The resonant interaction
does not disappear, because

\begin{equation}
(\boldsymbol A_u+\hat{\boldsymbol A}_r)^2=A_u^2+\hat A_r^2+2\boldsymbol A_u\cdot\hat{\boldsymbol A}_r.
\label{eq:3}
\end{equation}

The cross term mixes undulator and radiation harmonics. In a formulation retaining the transverse mechanical
momentum, the same interaction is represented through the undulator-induced $p_{\perp}$. The one-dimensional
theory is thus an effective longitudinal theory obtained after the transverse dynamics has supplied the
coupling. Its reduction and relation to quantum optics are discussed in \cite{ref25}.

We now specify the moving frame explicitly. Let $k_{r}=\omega_{r}/c>0$ be the laboratory wave number of the
forward radiation and $k_{u}=2\pi/\lambda_{u}>0$ the undulator spatial frequency. Select the static harmonic
with signed wave number $-k_{u}$. A boost along the electron beam with velocity $\beta_{B}c$ and Lorentz
factor $\gamma_{B}$ gives

\begin{equation}
\begin{aligned}
\omega'_r&=\gamma_B\omega_r(1-\beta_B),&k'_r&=\gamma_B k_r(1-\beta_B),\\
\omega'_u&=\gamma_B\beta_B c k_u,&k'_u&=-\gamma_B k_u.
\end{aligned}
\label{eq:4}
\end{equation}

The subscript $B$ labels the Bambini--Renieri boost; $\beta_{B}$ is its velocity divided by $c$, not the
undulator magnetic field. Primes denote BR-frame quantities. Choosing $\beta_{B}=k_{r}/(k_{r}+k_{u})$ makes
the two transformed frequencies equal. For a magnetic undulator, however, the transformed static harmonic
remains spacelike: $(\omega_{u}^{\prime}/c)^{2}-(k_{u}^{\prime})^{2}=-k_{u}^{2}$. It is therefore not an exact
free electromagnetic wave. In the ultrarelativistic boost limit it is approximately counterpropagating with
the radiation, so that

\begin{equation}
\omega'_u=\omega'_r\equiv\omega',\qquad k'_u\simeq-k'_r,\qquad |k'_u|\simeq k'_r\equiv k'.
\label{eq:5}
\end{equation}

The slow longitudinal electron motion in this frame motivates the BR reduction. For an optical undulator, both
modes are propagating waves and equal frequencies imply equal and opposite wave vectors exactly within the
two-wave model.

The positive recoil magnitude associated with creation of a radiation photon and removal of one pump quantum
is

\begin{equation}
q'=\hbar(k'_r-k'_u)=\frac{\hbar\omega'}{c}\left(1+\frac1{\beta_B}\right)\simeq2\hbar k'.
\label{eq:6}
\end{equation}

The middle expression applies to the static magnetic harmonic of Eq. (4); the last expression is its
ultrarelativistic approximation. For equal-frequency optical waves, $q^{\prime}=2\hbar k^{\prime}$ exactly.
Thus the customary notation $q=2\hbar k$ is the same recoil relation after primes have been suppressed, not an
additional momentum transfer. In the laboratory the corresponding spatial transfer is
$q_{\mathrm{lab}}=\hbar(k_{r}+$ $k_{u})$. These expressions refer to different frames; their transformation
also involves the exchanged energy. For clarity, let $\delta E$ and $\delta p_{z}$ denote the energy and
longitudinal momentum delivered to the fields in an exchange. Their boost from the laboratory to the BR frame
is

\begin{equation*}
\delta E'=\gamma_B(\delta E-\beta_B c\delta p_z),\qquad\delta p'_z=\gamma_B(\delta p_z-\beta_B\delta E/c).
\end{equation*}

For magnetic-undulator emission, $\delta E=\hbar\omega_{r}$ and $\delta p_{z}=\hbar(k_{r}+k_{u})$ in the
selected harmonic. Thus $q^{\prime}=\delta p_{z}^{\prime}$; choosing $\beta_{B}=k_{r}/(k_{r}+k_{u})$ gives
$\delta E^{\prime}=0$. This is the equal-frequency exchange in the BR description. Momentum transfer therefore
transforms as part of a four-vector, not as an independent scalar.

The physical meaning of an undulator pseudophoton can now be stated. A static laboratory magnet is not a gas
of real photons. In a rapidly moving frame its transverse magnetic field is accompanied by an electric field
and becomes a time-dependent periodic drive. The Weizsäcker--Williams equivalent-photon viewpoint represents
this drive by approximately wave-like Fourier components \cite{ref26}. In the present application its
usefulness rests on the near equality of transformed electric and magnetic field amplitudes when
$\beta_{B}\simeq1$, while the nonzero spacelike invariant noted above remains. It supplies a wave
representation of the interaction, not a literal quantization of the magnet as freely propagating radiation.

An equivalent occupation can be estimated only after specifying the harmonic and its normalization volume. For
a planar laboratory field $B_{y}=B_{u}\cos (k_{u}z)$, the transformed amplitudes obey $|B_{y}^{\prime}|=$
$\gamma_{B}|B_{y}|$ and $|E_{x}^{\prime}|=\gamma_{B}\beta_{B}c|B_{y}|$. The period-averaged field energy
density is consequently

\begin{equation}
\overline{u'}=\frac{\epsilon_0}{2}\overline{E'^2}+\frac1{2\mu_0}\overline{B'^2}=\frac{\gamma_B^2(1+\beta_B^2)B_u^2}{4\mu_0}.
\label{eq:7}
\end{equation}

Here $\epsilon_{0}$ and $\mu_{0}$ are the vacuum permittivity and permeability. Assigning energy
$\hbar\omega_{u}^{\prime}$ to each equivalent quantum in a BR-frame effective mode volume
$V_{\mathrm{eff}}^{\prime}$ gives the normalization estimate

\begin{equation}
\overline{\nu_u}^{\,\rm eq}=\frac{\overline{u'}V'_{\rm eff}}{\hbar\omega'_u}=\frac{\gamma_B(1+\beta_B^2)B_u^2 V'_{\rm eff}}{4\mu_0\hbar\beta_Bck_u}.
\label{eq:8}
\end{equation}

This is an equivalent-mode energy estimate for the chosen planar harmonic, not a frame-independent number of
photons stored in a static magnet. The volume, overlap and harmonic profile must be supplied for a numerical
value. A change of mode normalization changes both the assigned occupation and the single-quantum coupling;
the physical classical drive is their invariant product. For a true optical pump in a coherent state, by
contrast, $\overline{\nu_u}=|\alpha_{u}|^{2}$ is the actual mean occupation of the selected propagating mode.

The review article reported in Ref. \cite{ref27} interprets the undulator harmonic through equivalent massive
pseudophotons. The invariant above corresponds formally to the momentum scale $m_{\mathrm{ps}}=\hbar k_{u}/c$,
with a spacelike four-wavevector; it is not an on-shell photon of positive invariant mass. The static magnetic
field is supported by its physical sources. For the Maxwell--Proca massive-vector formulation and its
distinction from ordinary Maxwell electrodynamics, see Refs. \cite{ref28,ref29}. The Proca-like analogy here
organizes equivalent-quanta kinematics rather than implying a fundamental photon mass or a violation of
source-inclusive Maxwell equations. In particular, the spacelike undulator harmonic must not be identified
with an on-shell massive Proca particle.

For the following operator discussion we use BR coordinates and suppress primes:
$z=z^{\prime},p=p_{z}^{\prime}$ and $q=q^{\prime}$. Introducing $\hat{a}_{u}$ for the pump or its
equivalent-mode representation and $\hat{a}$ for radiation, the resonant two-mode interaction is

\begin{equation}
\hat H_{\rm int}^{(2)}=\hbar g_u\left(\hat a^\dagger\hat a_u\hat T_q+\hat a\hat a_u^\dagger\hat T_q^\dagger\right),\qquad \hat T_q=e^{-iq\hat z/\hbar}.
\label{eq:9}
\end{equation}

Here $g_{u}$ is the effective single-pump-quantum, single-radiation-quantum coupling, with dimensions of
angular frequency ($\mathrm{s}^{-1}$). It includes polarization, harmonic and spatial overlap. Its value is
fixed by the field normalization and the transverse reduction; it is not an independent universal constant.
For two equal-frequency transverse plane-wave modes in the nonrelativistic BR reduction, define their
vector-potential amplitudes per quantum by

\begin{equation}
A_{j,\rm zpf}=\sqrt{\frac{\hbar}{2\epsilon_0\omega' V'_j}},\quad g_u=\frac{e^2\xi A_{u,\rm zpf}A_{r,\rm zpf}}{m\hbar}=\frac{2\pi\alpha_{\rm fs}\hbar c\,\xi}{m\omega'\sqrt{V'_uV'_r}},\quad \alpha_{\rm fs}=\frac{e^2}{4\pi\epsilon_0\hbar c}.
\label{eq:10}
\end{equation}

Equivalently, with the reduced Compton wavelength $\bar{\lambda}_{C}=\hbar/(mc)$ and the common BR-frame
wavelength $\lambda^{\prime}=2\pi c/\omega^{\prime}$, the same coupling is

\begin{equation}
g_u=\frac{\alpha_{\rm fs}c\bar\lambda_C\lambda'\xi}{\sqrt{V'_uV'_r}}=\frac{\alpha_{\rm fs}c\lambda_C\lambda'\xi}{2\pi\sqrt{V'_uV'_r}},\qquad \lambda_C=\frac{h}{mc}=2\pi\bar\lambda_C.
\label{eq:11}
\end{equation}

Both wavelengths and the volumes in this expression are dimensional SI quantities; $g_{u}$ is in
$\mathrm{s}^{-1}$. The factor of $2\pi$ depends only on whether the ordinary or reduced Compton wavelength is
used.

The dimensionless overlap $\xi$ includes the polarization scalar product and the selected Fourier harmonic;
the phase may be absorbed in the mode operators. This expression follows from the resonant part of
$e^{2}\boldsymbol{A}_{u}\cdot\boldsymbol{A}_{r}/m$. It assumes negligible transverse mass dressing; a dressed
reduction changes the effective coefficient accordingly. For a magnetic-undulator pseudomode, the wave-like
normalization is only the equivalent-mode approximation described above. A classical pump replaces
$g_{u}\hat{a}_{u}$ by $g_{u}\alpha_{u}$, so the measurable reduced coupling is $g=g_{u}|\alpha_{u}|$,
independent of the arbitrary pump-mode normalization.

The operators act on the tensor-product space

\begin{equation}
\mathcal H=L^2(\mathbb R)_e\otimes\mathcal F_r\otimes\mathcal F_u,\qquad [\hat z,\hat p]=i\hbar,\qquad [\hat p,\hat T_q]=-q\hat T_q.
\label{eq:12}
\end{equation}

The electron coordinate and momentum act on the first factor; the two bosonic modes act on their respective
Fock spaces. The unitary operator $\hat{T}_{q}$ translates electron momentum and commutes with the field
operators. Momentum kets are generalized eigenstates; physical wave packets are superpositions of them. The
elementary transition and its bosonic matrix-element factor are

\begin{equation}
|p,n,\nu_u\rangle\longrightarrow|p-q,n+1,\nu_u-1\rangle,\qquad\sqrt{n+1}\sqrt{\nu_u}.
\label{eq:13}
\end{equation}

Emission is therefore a joint transition of electron and field, not a photon creation event with an unchanged
electron.

The interaction possesses a dressed Schwinger $su(2)$ algebra, developed in the Bloch-like treatment
\cite{ref30} and reviewed in \cite{ref27}. Define

\begin{equation}
\hat J_+=\hat a^\dagger\hat a_u\hat T_q,\qquad\hat J_-=\hat a\hat a_u^\dagger\hat T_q^\dagger,\qquad\hat J_z=\tfrac12(\hat n_r-\hat n_u).
\label{eq:14}
\end{equation}

Unitarity of $\hat{T}_{q}$ and the bosonic commutators give

\begin{equation}
[\hat J_z,\hat J_\pm]=\pm\hat J_\pm,\qquad [\hat J_+,\hat J_-]=2\hat J_z,\qquad\hat H_{\rm int}^{(2)}=\hbar g_u(\hat J_++\hat J_-).
\label{eq:15}
\end{equation}

The total mode occupation $\hat{N}=\hat{n}_{r}+\hat{n}_{u}$ and the combined recoil label
$\hat{P}=\hat{p}+q\hat{n}_{r}$ commute with this interaction. At fixed $N=n_{0}+\nu_{u0}$ and
$P=p_{0}+qn_{0}$, the coupled states are

\begin{equation}
|p_0-\ell q,n_0+\ell,\nu_{u0}-\ell\rangle,\qquad-n_0\le\ell\le\nu_{u0}.
\label{eq:16}
\end{equation}

This is a finite ladder with $N+1$ states, carrying spin $j=N/2$. Electron wave packets involve a continuum of
such conserved$-P$ sectors. The algebra thus organizes joint recoil and mode-exchange states; it does not
reduce the electron's full Hilbert space to a single spin.

The kinetic term is essential to their dynamics. As emphasized in \cite{ref27}, the full electron--field
algebra also contains momentum translations; the full Hamiltonian is not simply a spin in a prescribed
magnetic field. A Bloch-like representation alone therefore does not remove the kinetic recoil contribution.
For a longitudinal dispersion $E(p)$,

\begin{equation}
[E(\hat p),\hat J_+]|p,n,\nu_u\rangle=[E(p-q)-E(p)]\hat J_+|p,n,\nu_u\rangle.
\label{eq:17}
\end{equation}

It generally does not commute with the interaction. The free two-mode field energy additionally contributes
$\hbar(\omega_{r}^{\prime}-\omega_{u}^{\prime})$ to the energy difference of each exchange. The nonlinear
dependence of $E(p)$ on momentum generates recoil-dependent phases, so closure of the interaction operators as
$su(2)$ does not imply that the full Hamiltonian is linear in those generators. Deriving these phases is the
purpose of the next section.

For a macroscopic prescribed pump we replace its amplitude by a classical value,

\begin{equation}
\hat a_u\longrightarrow\sqrt{\overline{\nu_u}}e^{i\phi_u},\qquad g=g_u\sqrt{\overline{\nu_u}},\qquad N_{\rm exchanged}\ll\overline{\nu_u}.
\label{eq:18}
\end{equation}

For a static magnet, $\overline{\nu_u}$ has the equivalent-mode meaning of Eq. (8); only the resulting
classical drive is physical. For an optical pump the replacement assumes a large coherent amplitude with
negligible depletion. Its phase can be absorbed into the radiation operator. This assumption removes the
finite two-mode exchange symmetry and the dynamical pump occupation. It is a controlled large-pump reduction
of the interaction, not an algebra-preserving substitution. The remaining dressed radiation operators
$\hat{b}^{\dagger}=\hat{a}^{\dagger}\hat{T}_{q}$ and $\hat{b}=\hat{a}\hat{T}_{q}^{\dagger}$ obey
$[\hat{b},\hat{b}^{\dagger}]=1$: an oscillator algebra replaces the finite $su(2)$ ladder.

After the transverse reduction, resonant-harmonic selection and undepleted-pump approximation, the one-mode
Hamiltonian is

\begin{equation}
\hat H_{\rm eff}=E(\hat p)+\hbar\omega\hat a^\dagger\hat a+\hbar g\left(\hat a^\dagger\hat T_q+\hat a\hat T_q^\dagger\right).
\label{eq:19}
\end{equation}

The dimensional coupling $g$ has SI units $\mathrm{s}^{-1}$. The meaning of $\omega$ must be specified along
with the frame and energy reference. In a static laboratory-undulator description it is the radiation
frequency $\omega_{r}$, with $q=q_{\mathrm{lab}}$. In the BR two-mode description, the free field energy can
be written $\hbar\omega_{u}^{\prime}\hat{N}+\hbar(\omega_{r}^{\prime}-\omega_{u}^{\prime})\hat{n}_{r}$.
Removing the conserved first term, or equivalently retaining the pump phase in a rotating representation
before the classical replacement, gives $\omega=\omega_{r}^{\prime}-\omega_{u}^{\prime}$. At equal-frequency
BR resonance this coefficient vanishes, although the physical optical frequency $\omega^{\prime}$ does not.
Treating $\omega^{\prime}$ as the remaining photon-number coefficient would incorrectly omit the pump energy
from each transition.

Equation (19) is consequently a common form for explicitly specified effective descriptions. All quantities in
any calculation must use the same frame and energy reference. Below the generic dispersion
$E(p)=\sqrt{m^2c^4+c^2p^2}$ displays the longitudinal recoil construction; for a slow longitudinal electron in
the BR frame it reduces to $mc^{2}+p^{2}/(2m)$. If transverse averaging changes the effective dispersion, its
derivatives must be evaluated from that dispersion. A laboratory relativistic factor must not be inserted
independently into a BR kinetic term.

The omitted physics comprises off-resonant and counter-rotating harmonics, higher undulator harmonics, and
transverse dynamics already averaged out. Nonresonant energy shifts must be included in the adopted dispersion
or energy reference where relevant. These omissions define the model. Finally,

\begin{equation}
\hat T_q|p\rangle=|p-q\rangle
\label{eq:20}
\end{equation}

identifies the reduced emission event as $|p,n\rangle\to|p-q,n+1\rangle$. Its momentum translation and
kinetic-energy change inherit their meaning from the full undulator--radiation exchange.

\section{Interaction picture, relativistic recoil and the prescribed-current limit}

The interaction picture separates the free electron and radiation evolution from their coupling. Starting from
Eq. (19), write

\begin{equation}
\hat H=\hat H_0+\hat H_{\rm int},\qquad\hat H_0=E(\hat p)+\hat H_r,\qquad\hat H_r=\hbar\omega\hat a^\dagger\hat a,
\label{eq:21}
\end{equation}

with

\begin{equation}
E(p)=\sqrt{m^2c^4+c^2p^2},\qquad\hat H_{\rm int}=\hbar g(\hat a^\dagger\hat T_q+\hat a\hat T_q^\dagger),\qquad\hat T_q=e^{-iq\hat z/\hbar}.
\label{eq:22}
\end{equation}

Here $E(p)$ denotes the longitudinal dispersion in the adopted frame and reduction. The displayed square root
is the relativistic one-dimensional reference dispersion, not the slow-electron BR kinetic approximation. To
reconcile this expression with the results obtained in Ref. \cite{ref27}, we use $E_{\mathrm{BR}}(p)=$
$mc^{2}+p^{2}/(2m)$. The additive rest energy is the same before and after an exchange and therefore cancels
when their energy difference is formed. The subsequent ordered-operator derivation applies to either choice.
The wavevector transfer $q=q^{\prime}$ and time are BR quantities when using the BR Hamiltonian; laboratory
quantities are used only in a separately stated laboratory evaluation.

Following the standard interaction-picture construction \cite{ref31}, for any Schrödinger-picture operator,

\begin{equation}
\hat O_I(t)=e^{i\hat H_0t/\hbar}\hat O e^{-i\hat H_0t/\hbar}.
\label{eq:23}
\end{equation}

The electron and field operators act in different Hilbert spaces, so

\begin{equation}
[E(\hat p),\hat H_r]=0,\qquad e^{-i\hat H_0t/\hbar}=e^{-iE(\hat p)t/\hbar}e^{-i\hat H_rt/\hbar}.
\label{eq:24}
\end{equation}

Consequently,

\begin{equation}
\hat a_I(t)=\hat a e^{-i\omega t},\qquad\hat a_I^\dagger(t)=\hat a^\dagger e^{i\omega t}.
\label{eq:25}
\end{equation}

The electron translation requires more care. Its exact interaction-picture form is

\begin{equation}
\hat T_{q,I}(t)=e^{iE(\hat p)t/\hbar}e^{-iq\hat z/\hbar}e^{-iE(\hat p)t/\hbar}.
\label{eq:26}
\end{equation}

Acting on a momentum eigenstate gives successively

\begin{equation}
\begin{aligned}
\hat T_{q,I}(t)|p\rangle&=e^{iE(\hat p)t/\hbar}e^{-iq\hat z/\hbar}e^{-iE(p)t/\hbar}|p\rangle\\
&=e^{i[E(p-q)-E(p)]t/\hbar}|p-q\rangle.
\end{aligned}
\label{eq:27}
\end{equation}

Thus the free transformation itself supplies the energy difference. Including the photon creation phase leads
to

\begin{equation}
\Delta E(p,q)=E(p-q)-E(p)+\hbar\omega,\qquad\Delta(p,q)=\frac{\Delta E(p,q)}{\hbar}.
\label{eq:28}
\end{equation}

Here $\Delta E$ is an energy mismatch, measured in joules, while $\Delta$ is an angular-frequency mismatch,
measured in $\mathrm{s}^{-1}$. This convention will also be used in spectral kernels and dephasing estimates.
Detuning is not inserted as an unrelated phenomenological term: it originates in the free evolution.

Before any source approximation, the exact interaction-picture Hamiltonian of this reduced model is

\begin{equation}
\begin{aligned}
\hat H_I(t)&=\hbar g\hat a^\dagger e^{i\omega t}e^{iE(\hat p)t/\hbar}e^{-iq\hat z/\hbar}e^{-iE(\hat p)t/\hbar}\\
&\quad+\hbar g\hat a e^{-i\omega t}e^{iE(\hat p)t/\hbar}e^{iq\hat z/\hbar}e^{-iE(\hat p)t/\hbar}.
\end{aligned}
\label{eq:29}
\end{equation}

Keeping this ordering avoids treating a quantum electron phase prematurely as a classical trajectory. Under
the free electron Hamiltonian,

\begin{equation}
\frac{d\hat z_I}{dt}=\frac{i}{\hbar}[E(\hat p),\hat z_I]=\hat v(\hat p),\qquad\hat v(\hat p)=\frac{c^2\hat p}{E(\hat p)},\qquad\hat z_I(t)=\hat z+\hat v(\hat p)t.
\label{eq:30}
\end{equation}

However,

\begin{equation}
[\hat z,\hat v(\hat p)]=i\hbar\frac{\partial v}{\partial p}\ne0.
\label{eq:31}
\end{equation}

The exponential of $\hat{z}_{I}(t)$ therefore cannot be factorized as if position and velocity were commuting
numbers. The momentum-state calculation in Eq. (27) resolves this ordering issue directly.

The recoil content follows from the energy expansion

\begin{equation}
E(p-q)=E(p)-qE'(p)+\frac{q^2}{2}E''(p)-\frac{q^3}{6}E'''(p)+\cdots.
\label{eq:32}
\end{equation}

For the relativistic dispersion,

\begin{equation}
E'(p)=v,\qquad E''(p)=\frac{1}{m\gamma^3},\qquad\gamma=\frac{E(p)}{mc^2}.
\label{eq:33}
\end{equation}

Thus

\begin{equation}
\Delta(p,q)=\omega-\frac{qv}{\hbar}+\frac{q^2}{2\hbar m\gamma^3}+O(q^3/\hbar).
\label{eq:34}
\end{equation}

The first two terms give ordinary resonance mismatch; the displayed quadratic contribution is the leading
longitudinal recoil correction. The factor $m\gamma^{3}$ is the longitudinal effective mass for this specified
dispersion. For $|p|\ll mc$, expand the reference dispersion as $E(p)=mc^{2}+p^{2}/(2m)-$
$p^{4}/(8m^{3}c^{2})+\cdots$. Differentiating twice gives
$E^{\prime\prime}(p)=1/m-3p^{2}/(2m^{3}c^{2})+\cdots$. Thus $E_{\mathrm{BR}}^{\prime\prime}=1/m$ is the
leading nonrelativistic term, with relative corrections of order $p^{2}/(m^{2}c^{2})$, not an exact identity
for arbitrary momentum. If transverse quiver has been absorbed in an effective rest mass, that consistently
defined mass replaces $m$ throughout the reduced dispersion; no laboratory $\gamma^{3}$ is inserted
independently. The relativistic curvature is frame-dependent, not a scalar. In one longitudinal dimension,
$p^{\prime}=\gamma_{B}(p-\beta_{B}E/c),E^{\prime}=\gamma_{B}(E-\beta_{B}cp)$ and
$v^{\prime}=(v-\beta_{B}c)/(1-\beta_{B}v/c)$. Differentiating along the electron dispersion gives

\begin{equation*}
\frac{dv'}{dp'}=\frac{dv/dp}{\gamma_B^3(1-\beta_B v/c)^3}=\frac1{m[\gamma_B\gamma(1-\beta_B v/c)]^3}=\frac1{m\gamma'^3}.
\end{equation*}

For an electron slow in that frame, $\gamma^{\prime}\simeq1$ and the curvature becomes $1/m$ to the same
nonrelativistic accuracy. This argument concerns the one-dimensional reference dispersion; transverse
reduction must still be treated consistently.

For a helical undulator the transverse quiver magnitude is constant and can be incorporated in a consistent
effective-mass reduction; its magnitude must still justify any nonrelativistic expansion. For a planar
undulator the transverse kinetic contribution oscillates, so a boost alone does not give the same strictly
stationary one-dimensional nonrelativistic problem: transverse averaging and harmonic selection are additional
approximations.

After $\ell$ recoil events the electron momentum is $p_{0}-\ell q$, and the corresponding diagonal energy
difference is

\begin{equation}
\Delta E_\ell=E(p_0-\ell q)-E(p_0)+\ell\hbar\omega=\ell(\hbar\omega-qv_0)+\frac{\ell^2q^2}{2m\gamma_0^3}+\cdots.
\label{eq:35}
\end{equation}

As shown below, normalization gives a term linear in the exchanged-photon index and a quadratic recoil
curvature. Their distinct roles will be analyzed through neighboring-state mismatches in the Raman--Nath
equation.

The limit introduced next changes the dynamical problem: it produces radiation from an external source, not
self-consistent FEL amplification. For uniform prescribed motion the emission and absorption kernels have the
same mismatch, hence their stimulated difference is zero. In the early literature on quantum FEL dynamics
\cite{ref27}, the corresponding recoil-free seeded result is called gainless. This refers to net stimulated
amplification, not to the absence of photons emitted into an empty mode.

Only after this derivation do we introduce a prescribed current by replacing $\hat{z}_{I}(t)$ with an
externally specified classical trajectory $z_{c}(t)$. This is a physical approximation, not a change of
representation. The electron no longer responds dynamically to the field: recoil back-action, electron--field
entanglement and field-induced trajectory modification are removed. The Hamiltonian becomes

\begin{equation}
\hat H_{\rm src}^{(I)}(t)=\hbar g\left[\hat a^\dagger e^{i\omega t-iqz_c(t)/\hbar}+\hat a e^{-i\omega t+iqz_c(t)/\hbar}\right].
\label{eq:36}
\end{equation}

Define the complex driving coefficient $s(t)$ by

\begin{equation}
s(t)=-ig e^{i\omega t-iqz_c(t)/\hbar},\qquad\hat H_{\rm src}^{(I)}=i\hbar[s(t)\hat a^\dagger-s^*(t)\hat a].
\label{eq:37}
\end{equation}

The evolution is a displacement with a scalar ordering phase,

\begin{equation}
\hat U_I(t)=e^{i\Phi(t)}\hat D[\beta(t)],\quad\beta(t)=\int_0^t s(t')dt',\quad\hat D(\beta)=e^{\beta\hat a^\dagger-\beta^*\hat a}.
\label{eq:38}
\end{equation}

An initially empty mode therefore becomes a coherent state and has Poisson probabilities,

\begin{equation}
|0\rangle\longrightarrow e^{i\Phi}|\beta\rangle,\qquad P_n=e^{-\mu}\frac{\mu^n}{n!},\qquad\mu=|\beta|^2.
\label{eq:39}
\end{equation}

This is the prescribed-current reference problem. Setting the operator commutator in Eq. (31) to zero removes
its quantum ordering correction, but the approximation also removes the electron's dynamical response. It is
therefore not a general quantum-to-classical limit. A classical electron can recoil and exchange energy with a
field when its trajectory is solved self-consistently. In fact, as shown in Ref. \cite{ref27}, Eqs.
(7.61)--(7.66), a quantized field driven by such a semiclassical trajectory can retain coherent-state
statistics while reproducing classical gain. Coherence preservation in that approximation must not be confused
with the gainless straight-trajectory limit. It is not the self-consistent classical FEL, whose electron
phases and momenta remain dynamical and can develop modulation, bunching and collective gain.

Restoring the quantum electron replaces the specified phase by momentum-dependent energy differences.
Successive recoil states acquire different mismatches and the exact displacement structure is generally lost.
The Raman--Nath equations used below are coupled momentum-sideband equations: they describe how a periodic
electron--field interaction transfers probability between recoil orders, with the detuning and kinetic recoil
phases retained. The terminology recalls diffraction by a periodic grating. Here it does not mean that the
kinetic recoil term has already been discarded; that would be an additional short-interaction or recoilless
approximation.

\section{Vacuum fluctuations and spontaneous emission}

The prescribed-current solution has a useful connection with the coherent radiation structure of the
Bloch--Nordsieck soft-photon description \cite{ref32}. A prescribed charge current couples linearly to the
creation and annihilation operators; an initially empty mode is displaced into a coherent state. This supplies
a transparent reference problem with a classical external source and a quantum radiation field.

The Bloch--Nordsieck cancellation statement is more specific: infrared-finite inclusive probabilities require
the appropriate combination of real and virtual soft contributions. A single-mode FEL displacement calculation
illustrates the coherent-source structure, but is not a derivation of that cancellation theorem. Nor does it
justify discarding recoil when recoil is the observable of interest. The interaction-picture and phase-space
organization discussed later offers a complementary perspective \cite{ref33}.

For a dynamical electron the elementary emission and absorption matrix elements have the bosonic factors

\begin{equation}
|M_{\rm em}|^2\propto(n+1)|\langle p-q|\hat T_q|p\rangle|^2,\qquad|M_{\rm abs}|^2\propto n.
\label{eq:40}
\end{equation}

The extra unity follows from $\hat{a}\hat{a}^{\dagger}=\hat{a}^{\dagger}\hat{a}+1$. It is the vacuum
contribution to emission into an allowed empty mode, not a pre-existing real photon in the vacuum. It arises
from the nonzero vacuum matrix element of the creation operator; it is the transition probability that is
nonzero. Electron kinematics, mode structure, undulator coupling and interaction duration still determine the
probability.

Using the frequency mismatch $\Delta$ defined in Eq. (28), one may write schematic emission and absorption
kernels as

\begin{equation}
W_{\rm em}=\Gamma_+(p,\Delta)(n+1),\qquad W_{\rm abs}=\Gamma_-(p,\Delta)n.
\label{eq:41}
\end{equation}

For constant coupling over a physical duration $T$, define the per-bosonic-factor probabilities

\begin{equation}
\Gamma_+(p,T)=g^2T^2\sinc^2[\Delta_+(p)T/2],\qquad\Gamma_-(p,T)=g^2T^2\sinc^2[\Delta_-(p)T/2],
\label{eq:42}
\end{equation}

where $\Delta_{+}=\omega+[E(p-q)-E(p)]/\hbar$ and $\Delta_{-}=[E(p+q)-E(p)]/\hbar-\omega$. The even $\sinc$
makes either overall mismatch sign equivalent. Here $W_{\rm em,abs}$ in Eq. (41) denote probabilities, not
rates; these kernels retain $p,q$ and $T$, with mismatch dependence explicit. A shaped interaction replaces
$gT\sinc (\Delta T/2)$ by the appropriate time integral.

A finite-time probability is not a constant rate. Dividing by $T$ gives a duration-dependent average rate, not
a memoryless (Markov) population-evolution law. A time-independent rate requires an additional continuum or
reservoir limit and a coarse-graining interval longer than correlation times but shorter than the
population-evolution time. A selected coherent mode can instead display reversible exchange and must retain
its amplitudes. Explicit dimensionless finite-interaction kernels are derived in the next section. This keeps
the spontaneous factor separate from the dynamical line shape. Further comments on accumulated recoil and the
prescribed-current approximation are given in the concluding section.

Photon-number statistics and optical coherence must also be distinguished. An electron in a sharp momentum
state can become entangled with the radiation through distinct recoil orders. Tracing over distinguishable
electron states may destroy optical off-diagonal coherences, even when the resulting photon probabilities
happen to be Poissonian. The coherent state in Eq. (39) relies on the prescribed-current approximation; it is
not a conclusion about every recoilless electron--field preparation.

\section{Raman--Nath equation and local recoil expansion}

We first solve an initial-value problem for a specified input state: an electron with momentum $p_{0}$ and a
radiation mode containing exactly $n$ photons. According to the kinematics described above, the joint
electron--field state can be expanded as

\begin{equation}
|\Psi(t)\rangle=\sum_{\ell\ge-n}C_\ell(t)|p_0-\ell q,n+\ell\rangle.
\label{eq:43}
\end{equation}

The initial condition for this preparation is

\begin{equation}
C_\ell(0)=\delta_{\ell0},\qquad|\Psi(0)\rangle=|p_0\rangle\otimes|n\rangle.
\label{eq:44}
\end{equation}

It fixes an initially unexchanged electron--photon state. A sharp momentum ket is an idealized, delocalized
state, understood through box normalization or as the narrow-packet limit; it is not a general incoming
electron state. Experimentally that state results from cathode emission, extraction, acceleration, bending and
transport magnets, apertures and fluctuations in those stages. ``Preparation'' below is shorthand for the
resulting measured density operator at the interaction entrance, not an assumption that an ideal state can
simply be prescribed. Likewise, a photon number state is a legitimate conditional quantum state, but for a
phase-defined injected laser a coherent state is usually the more appropriate input. The vacuum used below is
both the number state $|0\rangle$ and the coherent state with zero amplitude. Choosing it removes the optical
preparation ambiguity, while leaving the electron preparation to be specified separately.

A set of conditional ladder solutions can also construct a general product input. Let $\phi(p)$ be a
normalized electron momentum amplitude and $c_{n}$ the optical number-state amplitudes. Then

\begin{equation*}
|\Psi(t)\rangle=\sum_{n=0}^\infty c_n\int dp\,\phi(p)\sum_{\ell\ge-n} A_\ell^{(n)}(p,t)|p-\ell q,n+\ell\rangle,
\end{equation*}

\begin{equation}
A_\ell^{(n)}(p,t)=e^{-i[E(p)+n\hbar\omega]t/\hbar}C_\ell^{(n)}(p,t/T_0),\qquad C_\ell^{(n)}(p,0)=\delta_{\ell0}.
\label{eq:45}
\end{equation}

Here $\int|\phi(p)|^{2}dp=1$ and $\sum_{n}|c_{n}|^{2}=1$. For a coherent input
$|\alpha_{0}\rangle,c_{n}=e^{-|\alpha_0|^2/2}\alpha_{0}^{n}/\sqrt{n!}$. Each conditional sector obeys its own
ladder equation, with the detuning and recoil curvature evaluated at its initial momentum. The factor
multiplying $C$ restores the common free phase removed within that sector; across a wave packet this phase
depends on $p$ and cannot be discarded. Coherent optical sectors must be superposed at the amplitude level,
including their relative phases, before tracing over the electron. Averaging number-state probabilities alone
would instead describe a phase-randomized optical preparation. Mixed electron or optical inputs are treated by
evolving their density matrices linearly. The consequences of a finite electron packet are developed in
Section VI before introducing its Wigner representation.

The diagonal mismatch is already given by Eq. (35); it need not be expanded a second time. To fix every
normalization, choose a positive reference interaction time $T_{0}$ in seconds. It can be chosen as the
transit duration through the undulator. For an interaction length $L_{u}=N_{u}\lambda_{u}$ in the laboratory,
a central electron with velocity $v_{0}$ has $T_{\rm tr,lab}=L_{u}/v_{0}$. The duration along the same orbit
in the BR frame is $T_{\rm tr,BR}=\gamma_{B}(T_{\rm tr,lab}-\beta_{B}L_{u}/c)$. Because the undulator
boundaries move in the BR frame, the transit is $T_{\rm
tr,BR}=(L_{u}/\gamma_{B})/(v_{0}^{\prime}+\beta_{B}c)$, where
$v_{0}^{\prime}=(v_{0}-\beta_{B}c)/(1-\beta_{B}v_{0}/c)$. Writing $L_{u,\rm BR}/v_{0,\rm BR}$ is correct only
if the denominator denotes the electron velocity relative to the moving undulator boundaries, not its BR
velocity alone. This reproduces the Lorentz-transformed duration above. If $T_{0}=T_{\mathrm{tr}}$ in the
adopted frame, the exit occurs at $\tau=1$. Alternatively $T_{0}$ is a free positive normalization scale, in
which case the physical exit is at $\tau_{\mathrm{tr}}=T_{\mathrm{tr}}/T_{0}$; changing $T_{0}$ changes the
dimensionless parameters but not the dynamics at fixed physical time. A packet narrow enough that its
transit-time variation is negligible can use the central-orbit duration. Define

\begin{equation}
\tau=\frac{t}{T_0},\qquad\mathfrak g=gT_0,\qquad W=\left(\frac{qv_0}{\hbar}-\omega\right)T_0,\qquad\varepsilon=\frac{q^2T_0}{2\hbar m\gamma_0^3}.
\label{eq:46}
\end{equation}

For $q=\hbar Q$ with a geometry-defined spatial wave number $Q,W=(Qv_{0}-\omega)T_{0}$ is independent of
Planck's constant. By contrast $\varepsilon=\hbar Q^{2}E^{\prime\prime}T_{0}/2$ contains the recoil scale.
Thus $\tau,\mathfrak{g},W$ and $\varepsilon$ are dimensionless; $g$ retains its earlier dimensional meaning.
For a different effective dispersion, the last definition is
$\varepsilon=q^{2}E^{\prime\prime}(p_{0})T_{0}/(2\hbar)$. In the slow-electron BR dispersion
$E^{\prime\prime}=1/m$. No unspecified reference frequency or new coupling is needed.

After removal of the common initial free phase, the Raman--Nath ladder is

\begin{equation}
i\frac{dC_\ell}{d\tau}=(-W\ell+\varepsilon\ell^2)C_\ell+\mathfrak g\left[\sqrt{n+\ell+1}\,C_{\ell+1}+\sqrt{n+\ell}\,C_{\ell-1}\right].
\label{eq:47}
\end{equation}

The boundary $n+\ell\ge0$ follows from photon-number positivity. The square-root factors encode stimulated and
vacuum emission; the diagonal terms encode detuning and recoil. The Taylor reduction of the electron energy
remains controlled only while the occupied momentum shifts stay in its validity range. Higher recoil orders
are not automatically described accurately by the quadratic approximation.

For the conditional vacuum input $n=0$, initially sharp electron momentum and initial condition Eq. (44) used
in the following perturbative calculation, remove the diagonal phase exactly:

\begin{equation}
C_\ell(\tau)=e^{-i(-W\ell+\varepsilon\ell^2)\tau}D_\ell(\tau).
\label{eq:48}
\end{equation}

Substitution into Eq. (47) gives

\begin{equation}
\begin{aligned}
i\dot D_\ell&=\mathfrak g\sqrt{\ell+1}\,e^{iW\tau-i\varepsilon(2\ell+1)\tau}D_{\ell+1}\\
&\quad+\mathfrak g\sqrt\ell\,e^{-iW\tau+i\varepsilon(2\ell-1)\tau}D_{\ell-1}.
\end{aligned}
\label{eq:49}
\end{equation}

The dot denotes differentiation with respect to $\tau$. The neighboring diagonal differences are

\begin{equation}
d_{\ell+1}-d_\ell=-W+\varepsilon(2\ell+1),\quad d_\ell-d_{\ell-1}=-W+\varepsilon(2\ell-1),\quad d_\ell=-W\ell+\varepsilon\ell^2.
\label{eq:50}
\end{equation}

The local mismatch means the energy difference between two neighboring exchange states, divided by the
reference frequency $1/T_{0}$, including their detuning. This difference represents the local mismatch and
fixes the phase of that transition. The local mismatch, rather than the absolute diagonal energy
$\varepsilon\ell^{2}$, controls the transition phases. This is the physical reason for expanding the
transformed couplings.

We assume that

\begin{equation}
|\varepsilon(2\ell\pm1)\tau|\ll1
\label{eq:51}
\end{equation}

throughout the significantly populated ladder. Keeping $W$ nonzero and exact, the phase factors become

\begin{equation}
\begin{aligned}
e^{iW\tau-i\varepsilon(2\ell+1)\tau}&=e^{iW\tau}\left[1-i\varepsilon(2\ell+1)\tau-\frac{\varepsilon^2}{2}(2\ell+1)^2\tau^2\right]+O(\varepsilon^3),\\
e^{-iW\tau+i\varepsilon(2\ell-1)\tau}&=e^{-iW\tau}\left[1+i\varepsilon(2\ell-1)\tau-\frac{\varepsilon^2}{2}(2\ell-1)^2\tau^2\right]+O(\varepsilon^3).
\end{aligned}
\label{eq:52}
\end{equation}

Writing

\begin{equation}
D_\ell=D_\ell^{(0)}+\varepsilon D_\ell^{(1)}+\varepsilon^2D_\ell^{(2)}+O(\varepsilon^3),
\label{eq:53}
\end{equation}

introduce the exact recoilless, detuned ladder operator

\begin{equation}
(\mathcal H_0D)_\ell=\G\left(e^{iW\tau}\sqrt{\ell+1}\,D_{\ell+1}+e^{-iW\tau}\sqrt\ell\,D_{\ell-1}\right).
\label{eq:54}
\end{equation}

The various orders of the perturbative expansion are linked by the differential equations

\begin{equation}
i\dot D^{(0)}=\mathcal H_0D^{(0)},\quad i\dot D^{(1)}=\mathcal H_0D^{(1)}+S^{(1)},\quad i\dot D^{(2)}=\mathcal H_0D^{(2)}+S^{(2)},
\label{eq:55}
\end{equation}

where

\begin{equation}
\begin{aligned}
S_\ell^{(1)}&=-i\G\tau(2\ell+1)e^{iW\tau}\sqrt{\ell+1}\,D_{\ell+1}^{(0)}\\
&\quad+i\G\tau(2\ell-1)e^{-iW\tau}\sqrt\ell\,D_{\ell-1}^{(0)},
\end{aligned}
\label{eq:56}
\end{equation}

and

\begin{equation}
\begin{aligned}
S_\ell^{(2)}&=-i\G\tau(2\ell+1)e^{iW\tau}\sqrt{\ell+1}\,D_{\ell+1}^{(1)}+i\G\tau(2\ell-1)e^{-iW\tau}\sqrt\ell\,D_{\ell-1}^{(1)}\\
&\quad-\frac{\G\tau^2}{2}(2\ell+1)^2 e^{iW\tau}\sqrt{\ell+1}\,D_{\ell+1}^{(0)}-\frac{\G\tau^2}{2}(2\ell-1)^2 e^{-iW\tau}\sqrt\ell\,D_{\ell-1}^{(0)}.
\end{aligned}
\label{eq:57}
\end{equation}

If $U_{0}(\tau,s)$ is the propagator of Eq. (54), and the initial state has no recoil-dependent correction,
then

\begin{equation}
D^{(j)}(\tau)=-i\int_0^\tau U_0(\tau,s)S^{(j)}(s)\,ds,\qquad j=1,2.
\label{eq:58}
\end{equation}

Equation (58) is the Duhamel formula, or variation-of-constants solution \cite{ref34}, for
$i\partial_{\tau}D^{(j)}=$ $\mathcal H_0(\tau)D^{(j)}+S^{(j)}(\tau)$ with $D^{(j)}(0)=0$. Here $U_0(\tau,s)$
satisfies $i\partial_\tau U_0=\mathcal H_0(\tau)U_0$ and $U_0(s,s)=I$. Multiplying by $U_{0}(s,\tau)$ and
integrating yields the formula; differentiating it recovers the inhomogeneous equation and its initial
condition. It is an exact identity at each hierarchy order. Its iteration is the time-ordered Dyson
construction \cite{ref35,ref36}; only the subsequent finite-order truncation is approximate. The ``Duhamel
integrals'' below refer specifically to these source-propagation integrals.

These equations give an operational expansion: detuning is retained exactly, while recoil enters through local
mismatch and explicit source terms.

For one elementary neighboring transition, define the dimensionless frequency argument $\nu$ and its
finite-time integral

\begin{equation}
\mathcal A(\nu,\tau)=\int_0^\tau e^{i\nu s}ds=\tau e^{i\nu\tau/2}\sinc(\nu\tau/2),\quad \mathcal S_0(W,\tau)=\tau^2\sinc^2(W\tau/2).
\label{eq:59}
\end{equation}

We use $\sinc x=\sin x/x$ with $\sinc 0=1$. These forms are explicitly regular: $\mathcal{A}(0,\tau)=\tau$ and
$\mathcal{S}_{0}(0,\tau)=\tau^{2}$. Any denominator in the equivalent trigonometric expressions is therefore a
removable singularity, not a physical divergence. A local recoil shift
$\delta_{\ell}^{(\pm)}=\varepsilon(2\ell\pm1)$ yields

\begin{equation}
\mathcal S_0(W-\delta,\tau)=\mathcal S_0+\delta\mathcal G_1+\delta^2\mathcal G_2+O(\delta^3),
\label{eq:60}
\end{equation}

with the explicit functions

\begin{equation}
\mathcal G_1(W,\tau)=-\partial_W\mathcal S_0=\frac4{W^3}\left[1-\cos(W\tau)-\frac{W\tau}{2}\sin(W\tau)\right],
\label{eq:61}
\end{equation}

\begin{equation}
\mathcal G_2(W,\tau)=\frac12\partial_W^2\mathcal S_0=\frac{W^2\tau^2\cos(W\tau)-4W\tau\sin(W\tau)-6\cos(W\tau)+6}{W^4}.
\label{eq:62}
\end{equation}

The apparent inverse powers of $W$ in these derivatives also cancel. Expanding the entire line shape about
$W=0$ gives

\begin{equation}
\begin{aligned}
\mathcal S_0&=\tau^2-\frac{W^2\tau^4}{12}+\frac{W^4\tau^6}{360}+O(W^6\tau^8),\\
\mathcal G_1&=\frac{W\tau^4}{6}-\frac{W^3\tau^6}{90}+O(W^5\tau^8),\\
\mathcal G_2&=-\frac{\tau^4}{12}+\frac{W^2\tau^6}{60}+O(W^4\tau^8).
\end{aligned}
\label{eq:63}
\end{equation}

Thus $\mathcal{G}_{1}(0,\tau)=0$ and $\mathcal{G}_{2}(0,\tau)=-\tau^{4}/12$. These series, or the $\sinc$
representation and its derivatives, should be used numerically near zero detuning to avoid cancellation
errors. The amplitude hierarchy is itself regular there: its source terms contain only polynomials in $\tau$
and exponentials $e^{\pm iW\tau}$. On a finite ladder the Duhamel integrals have no zero-detuning singularity.
Their use on the infinite ladder additionally requires the population and phase control discussed in Section
VI.

At $\tau=1,\mathcal{G}_{1}(W,1)=-\partial_{W}\sinc ^{2}(W/2)$ is the usual antisymmetric low-gain FEL
function, apart from the conventional multiplicative gain coefficient. In the normalization of Ref.
\cite{ref37}, the space-charge-free small-signal gain is $G(W,0)=\pi g_{0}\mathcal{G}_{1}(W,1)$. The frequency
orientation must be matched when comparing signs. This identifies the physical meaning of the first derivative
without introducing a separate complex-response hierarchy.

More generally $\mathcal{G}_{j}=(-1)^{j}\partial_{W}^{j}\mathcal{S}_{0}/j!$. The first derivative is
antisymmetric in $W$, and the curvature correction is even. Locally each rung samples a shifted line shape;
the full probability still requires the coupled amplitude evolution in Eq. (55), rather than a separate
replacement of the mean in every probability.

For an initially sharp electron momentum and $n$ initial photons, first-order transition amplitudes give, to
second order in the coupling,

\begin{equation}
P_{\rm em}\simeq\G^2(n+1)\mathcal S_0(W-\varepsilon,\tau),\qquad P_{\rm abs}\simeq\G^2 n\mathcal S_0(W+\varepsilon,\tau).
\label{eq:64}
\end{equation}

The stimulated photon change is consequently

\begin{equation}
\Delta n_{\rm stim}\simeq n\G^2[\mathcal S_0(W-\varepsilon,\tau)-\mathcal S_0(W+\varepsilon,\tau)]=2n\G^2\varepsilon\mathcal G_1(W,\tau)+O(\varepsilon^3).
\label{eq:65}
\end{equation}

Equation (65) is evaluated to leading order $g^2$ in the transition probabilities. The displayed remainder
refers to the recoil expansion at that order; higher orders in the coupling are also neglected.

The remaining $\mathfrak{g}^{2}\mathcal{S}_{0}(W-\varepsilon,\tau)$ term is spontaneous feeding of the
selected mode. This explicitly connects the $n+1$ factor, recoil shifts and the derivative structure of the
small-signal gain \cite{ref1,ref7}. The displayed result is a weak-transition single-electron expression, not
a self-consistent high-gain growth law.

We checked the amplitude hierarchy directly against propagation of the time-independent Raman--Nath
Hamiltonian. For the vacuum initial condition, $\mathfrak{g}=0.7$ and $\tau=1$, let

\begin{equation}
\mathcal E_2=\left\|C_{\rm exact}-e^{-id_\ell\tau}\left(D^{(0)}+\varepsilon D^{(1)}+\varepsilon^2D^{(2)}\right)\right\|_2,\qquad d_\ell=-W\ell+\varepsilon\ell^2.
\label{eq:66}
\end{equation}

The phase multiplication in this expression is componentwise. Matrix-exponential propagation supplies
$C_{\mathrm{exact}}$; numerical integration of Eq. (55) supplies the three perturbative coefficients. The
results are

\begin{center}
\begin{tabular}{ccc}
$W$ & $\varepsilon$ & $\mathcal E_2$\\ \hline
0   & 0.04 & $1.0551\times10^{-3}$\\
0   & 0.02 & $1.3344\times10^{-4}$\\
0   & 0.01 & $1.6729\times10^{-5}$\\
0.8 & 0.04 & $9.5773\times10^{-4}$\\
0.8 & 0.02 & $1.2028\times10^{-4}$\\
0.8 & 0.01 & $1.5028\times10^{-5}$
\end{tabular}
\end{center}

Increasing the truncation from $0\le\ell\le35$ to $0\le\ell\le47$ changes the reported errors by less than
$10^{-12}$. The hierarchy was integrated with relative tolerance $2\times10^{-12}$ and absolute tolerance
$2\times10^{-14}$. Halving $\varepsilon$ reduces the error by approximately eight, consistent with an omitted
cubic amplitude correction, including at $W=0$. At $\varepsilon=0$ the numerical photon probabilities agree
with the analytic Poisson law to within $4\times10^{-15}$. Here the normalization residuals mean
$|\sum_{\ell}|D_{\ell}^{(0)}|^{2}-1|$,

$|\sum_{\ell}2\operatorname{Re} (D_{\ell}^{(0)\ast}D_{\ell}^{(1)})|$ and
$|\sum_{\ell}[|D_{\ell}^{(1)}|^{2}+2\operatorname{Re} (D_{\ell}^{(0)\ast}D_{\ell}^{(2)})]|$. All three are
below $5\times10^{-15}$ in these tests. These checks show that the terms of order $1,\varepsilon$ and
$\varepsilon^{2}$ in the norm have the required values. At nonzero $\varepsilon$, the truncated state can
still have a norm error from terms of order $\varepsilon^{3}$ and higher; that error must be checked
separately. These are checks of the conditional ladder calculation in this parameter range, not a claim of
uniform accuracy at arbitrary coupling, time or occupied recoil order.

\section{Photon statistics and FEL-induced energy spread}

The earlier many-particle analysis of Becker and McIver \cite{ref9} uses an interaction-picture recoil
expansion and treats photon fluctuations as well as mean gain. Their review \cite{ref10} places photon
statistics and cooperative effects at the centre of the quantum description. This historical connection is
important: non-Poissonian startup statistics are not by themselves a unique signature of a recoil-resolved
oscillator, and an observable prediction must retain the relevant preparation and beam averaging.

In the preceding calculation we treated the vacuum-input RN equation, $n=0$. Its $\sqrt{\ell+1}$ and
$\sqrt\ell$ factors originate from photon creation and annihilation. At zero recoil its solution has Poisson
probability amplitudes. Within a fixed initial photon-number sector, for a strongly populated mode with
$n\gg|\ell|$ over every populated exchange order, both square roots can instead be approximated by $\sqrt n$
and absorbed into the coupling $h=\mathfrak{g}\sqrt n$. The equation then becomes the shift RN equation
\cite{ref36}, related to the original ultrasound diffraction problem \cite{ref38}:

\begin{equation}
i\dot C_\ell=(-W\ell+\varepsilon\ell^2)C_\ell+h(C_{\ell+1}+C_{\ell-1}).
\label{eq:67}
\end{equation}

The lower photon boundary is remote in this approximation. At $\varepsilon=0$, for
$C_{\ell}(0)=\delta_{\ell0}$ and effectively integer $\ell$ of either sign,

\begin{equation}
C_\ell^{(0)}(\tau)=(-i)^\ell e^{i\ell W\tau/2}J_\ell[2h\tau\sinc(W\tau/2)].
\label{eq:68}
\end{equation}

Because $J_{-\ell}^{2}=J_{\ell}^{2}$, the fixed-number exchange probabilities are symmetric and
$\langle\ell\rangle=0$ in this recoil-free approximation. This is an exchange-order distribution referenced to
$|n\rangle$, not the complete photon-count distribution of a coherently seeded laser. The Poisson and Bessel
formulas do not identify two distinct physical recoil regimes. They are representations or approximations
within the same recoil-free dynamics, evaluated with different input-state specifications. For a coherent
seed, the number-sector amplitudes in Eq. (45) must be superposed with
$c_{n}=e^{-|\alpha_0|^2/2}\alpha_{0}^{n}/\sqrt{n!}$ before constructing the final density operator. Replacing
the coherent seed by $|n\rangle$, or by an incoherent average of number-state exchange probabilities, removes
information required for phase-sensitive observables.

The common coherent-state structure is explicit in the prescribed-current reference. Let $\beta$ be the
displacement generated by the source in the radiation interaction picture. Then

\begin{equation}
U_I=e^{i\Phi}D(\beta),\quad D(\beta)|\alpha_0\rangle=e^{(\beta\alpha_0^*-\beta^*\alpha_0)/2}|\alpha_0+\beta\rangle,\quad D(\beta)=e^{\beta\hat a^\dagger-\beta^*\hat a}.
\label{eq:69}
\end{equation}

Vacuum is the case $\alpha_{0}=0$; a nonzero coherent seed obeys the same displacement evolution. At a fixed
source and optical phase both give a Poisson law with mean $|\alpha_{0}+\beta|^{2}$. A Bessel exchange kernel
in a fixed-number approximation cannot be substituted for that law. Phase-dependent interference can change
the field energy for a coherent seed, so the vanishing mean exchange in the fixed-number calculation must not
be transferred indiscriminately to it.

When the electron is retained as a quantum degree of freedom, the zero-recoil displacement involves the
dressed operator $\hat{A}$, not simply the bare radiation operator. The reduced optical statistics can then
depend on electron coherence and tracing; a coherent optical seed alone does not ensure an unentangled
coherent output. The appropriate comparison therefore keeps the same Hamiltonian and the stated electron and
optical preparation. The point made in Ref. \cite{ref39} is more specific: treating the undulator pump as a
photon-number state produces antibunching even at zero electron recoil. Such a state does not represent the
coherent classical drive intended in that calculation. The coherent pump removes this artifact. This concerns
the undulator mode, not merely the choice of a radiation seed.

To display the result, retain both modes in Eq. (9), take zero initial radiation photons and zero recoil, and
let $m$ be the initial pump occupation. The exact spherical Raman--Nath solution in Ref. \cite{ref39} gives

\begin{equation}
P(\ell\mid m)=\binom m\ell\chi^\ell(1-\chi)^{m-\ell},\quad \chi=\frac{4\G_u^2}{4\G_u^2+W^2}\sin^2\left[\frac\tau2\sqrt{4\G_u^2+W^2}\right],\quad\G_u=g_uT_0.
\label{eq:70}
\end{equation}

Here $0\le\ell\le m$ counts transferred pump quanta and emitted radiation photons, and $\chi$ is the
single-quantum transfer probability. The phase of the amplitudes is needed for coherence observables but not
for this population law. A pump Fock state has

\begin{equation}
\langle\ell\rangle_m=m\chi,\quad\Var(\ell\mid m)=m\chi(1-\chi),\quad Q_M=\frac{\Var(\ell)-\langle\ell\rangle}{\langle\ell\rangle}=-\chi,
\label{eq:71}
\end{equation}

for nonzero mean. The symbol $Q_{M}$ denotes the Mandel parameter, distinct from cavity quality factor or
bunch charge. The negative value is present without recoil; it is inherited from the number-state pump.

For a coherent undulator input $|\beta_{u}\rangle$, let $\bar{m}=|\beta_{u}|^{2}$. Averaging the number
populations with the coherent-state weights yields

\begin{equation*}
P_\ell=\sum_{m=\ell}^\infty e^{-\bar m}\frac{\bar m^m}{m!}\binom m\ell\chi^\ell(1-\chi)^{m-\ell}=e^{-\chi\bar m}\frac{(\chi\bar m)^\ell}{\ell!},
\end{equation*}

\begin{equation}
\langle\ell\rangle=\chi\bar m,\qquad\Var(\ell)=\chi\bar m,\qquad Q_M=0.
\label{eq:72}
\end{equation}

This is the explicit correction to the artificial zero-recoil antibunching. The counting calculation depends
on the pump number weights; a phase-random state with those same Poisson weights gives the same populations
but not the same phase coherence. The coherent-pump amplitudes are required for the coherent Bloch-state
interpretation used in the reference. Poisson counts alone still do not prove a pure coherent reduced
radiation state after tracing out the electron. In the large-pump, weak-depletion limit,
$\chi\bar{m}\to\mathfrak{g}^{2}\mathcal{S}_{0}$ with $\mathfrak{g}=\mathfrak{g}_{u}\sqrt{\bar m}$, recovering
the one-mode Poisson reference below. Thus the appropriate recoil test first specifies the pump preparation
and then compares finite-recoil predictions with this corrected zero-recoil baseline.

Before proceeding, we underline the role of $\hat{A}=\hat{a}\hat{T}_{q}^{\dagger}$, the dressed annihilation
operator. At zero recoil the vacuum exchange-ladder state evolves into a coherent state in the Glauber sense
of this joint electron--field operator. Kinetic recoil breaks that coherence-preserving dynamical structure
and generally produces a state that is not an eigenstate of $\hat{A}$, together with non-Poissonian
probabilities. This is joint coherence; it does not by itself make the reduced optical state a pure coherent
state. The operator algebra $[\hat{A},\hat{A}^{\dagger}]=1$ remains valid even when recoil is present; it is
the evolution's coherent-state preservation that fails.

The diagonal transformation does not change probabilities:

\begin{equation}
P_\ell=|C_\ell|^2=|D_\ell|^2=P_\ell^{(0)}+\varepsilon P_\ell^{(1)}+\varepsilon^2 P_\ell^{(2)}+O(\varepsilon^3),
\label{eq:73}
\end{equation}

where

\begin{equation}
P_\ell^{(1)}=2\operatorname{Re}[D_\ell^{(0)*}D_\ell^{(1)}],\qquad P_\ell^{(2)}=|D_\ell^{(1)}|^2+2\operatorname{Re}[D_\ell^{(0)*}D_\ell^{(2)}].
\label{eq:74}
\end{equation}

At order $\varepsilon^{2}$, two contributions arise when the amplitude is squared: the product of the
first-order correction with its conjugate, and the interference of the zeroth- and second-order amplitudes.
Both must be retained for a consistent second-order probability.

For the conditional preparation $|p_{0}\rangle\otimes|0\rangle$ with $C_{\ell}(0)=\delta_{\ell0}$, at
$\varepsilon=0$ the vacuum ladder amplitudes have the displaced-oscillator form, yielding

\begin{equation}
\mu(W,\tau)=\G^2\mathcal S_0(W,\tau),\qquad P_\ell^{(0)}=e^{-\mu}\frac{\mu^\ell}{\ell!},\qquad F^{(0)}=1,\qquad g_0^{(2)}(0)=1.
\label{eq:75}
\end{equation}

Here ``conditional'' means calculated at a fixed initial electron momentum $p$ and specified initial photon
number, before averaging over the entrance beam distribution; it does not imply detector postselection. The
variance is explicitly

\begin{equation}
\Var(n)=\langle(n-\langle n\rangle)^2\rangle=\sum_n n^2P_n-\left(\sum_n nP_n\right)^2.
\label{eq:76}
\end{equation}

It is the squared rms photon-number fluctuation. For a nonzero mean photon number, $F=$ $\Var (n)/\langle
n\rangle$ and $g^{(2)}(0)=\langle n(n-1)\rangle/\langle n\rangle^{2}$. The coherent ladder-amplitude structure
does not prove that the physical reduced field is a coherent state: the joint state Eq. (43) still ties each
photon number to a distinct electron recoil. The prescribed-source state and the joint sharp-momentum state
are different preparations. For a finite electron packet and vacuum input, the conditional Poisson laws must
instead be integrated over the initial momentum density. Explicitly, $\langle\mu\rangle_{f_0}=\int
dp\,f_{0}(p)\mu(p)$ and $\Var _{f_0}(\mu)=\int dp\,f_{0}(p)[\mu(p)-\langle\mu\rangle_{f_0}]^{2}$. To derive
the variance, use the conditional Poisson identities $\langle n\rangle_{p}=\mu(p)$ and $\langle
n^{2}\rangle_{p}=\mu(p)^{2}+\mu(p)$. Averaging gives $\langle n\rangle=\langle\mu\rangle_{f_0}$ and $\langle
n^{2}\rangle=\langle\mu^{2}\rangle_{f_0}+\langle\mu\rangle_{f_0}$. Subtracting the square of the mean yields
the law of total variance:

\begin{equation}
\Var(n)=\langle\Var(n|p)\rangle_{f_0}+\Var_{f_0}(\langle n|p\rangle).
\label{eq:77}
\end{equation}

Consequently the complete zeroth-order photon variance is

\begin{equation}
\Var(n)=\langle\mu\rangle_{f_0}+\Var_{f_0}(\mu).
\label{eq:78}
\end{equation}

For the recoil-corrected vacuum probabilities, write $M_{r}=\sum_{\ell\ge0}\ell^{r}P_{\ell}$; then $\Var
(n)=M_{2}-M_{1}^{2}$. If

$M_{r}=\sum_{j=0}^{2}\varepsilon^{j}M_{r}^{(j)}$, its second-order consistent expression is

\begin{equation}
\begin{aligned}
\Var(n)&=M_2^{(0)}-(M_1^{(0)})^2+\varepsilon[M_2^{(1)}-2M_1^{(0)}M_1^{(1)}]\\
&\quad+\varepsilon^2[M_2^{(2)}-2M_1^{(0)}M_1^{(2)}-(M_1^{(1)})^2]+O(\varepsilon^3).
\end{aligned}
\label{eq:79}
\end{equation}

A mixture with momentum-dependent $\mu$ is generally not Poissonian. Its Fano factor is

\begin{equation}
F=1+\frac{\Var_{f_0}(\mu)}{\langle\mu\rangle_{f_0}}\ge1.
\label{eq:80}
\end{equation}

Thus an initial spread can alter photon statistics even when the conditional recoil parameter is zero.

The exact RN Hamiltonian is Hermitian and its evolution unitary. A finite recoil expansion constructed by the
Duhamel/Dyson iteration \cite{ref35,ref36} is not exactly unitary: normalization holds only through the
retained order, and the residual must be monitored together with boundary population and positivity.
Renormalizing a truncated probability distribution does not restore omitted physical terms. The numerical and
order-by-order checks reported above apply to the stated parameter range.

Recoil deforms the distribution through the index-dependent coupling phases, not through the removed diagonal
gauge phase itself. The expansion is not restricted by definition to small signal. Its control condition is
Eq. (51) on every appreciably occupied transition throughout the interval. We make ``appreciably occupied''
quantitative by choosing a discarded-probability tolerance $\delta_{\mathrm{tail}}$ and defining the Poisson
cutoff

\begin{equation}
L_{\delta_{\rm tail}}(\mu)=\min\left\{L\in\mathbb N_0:\sum_{\ell=0}^L e^{-\mu}\frac{\mu^\ell}{\ell!}\ge1-\delta_{\rm tail}\right\}.
\label{eq:81}
\end{equation}

For example, $\delta_{\mathrm{tail}}=10^{-6}$ gives $L_{\delta_{\rm tail}}=4,9,28,151$ for $\mu=0.1,1,10,100$,
respectively. The mean relevant to a finite interval is $\mu_{\mathrm{\max }}=\max _{0\le
s\le\tau_{\max}}\mathfrak{g}^{2}\mathcal{S}_{0}(W,s)$, not necessarily its final value. A conservative phase
criterion for the retained ladder and its upper neighboring transition is

\begin{equation}
|\varepsilon|\big[2L_{\delta_{\rm tail}}(\mu_{\max})+1\big]\tau_{\max}\ll1.
\label{eq:82}
\end{equation}

This replaces an unspecified tail-selection constant by a stated probability tolerance. It is a control
estimate based on the recoilless distribution; an actual calculation must also monitor recoil-induced
population outside the chosen cutoff and convergence when the cutoff is increased. A norm or observable error
test, such as Eq. (66), gives a separate quantitative accuracy check. For an initial packet the criterion must
hold over the retained range of initial momenta as well.

Small signal often keeps the populated ladder short, but is not the mathematical definition of the recoil
expansion. In stronger interaction, even a small $|\varepsilon|$ can accumulate a significant phase as
populated orders move and broaden. In that case retain the factors $e^{iW\tau-i\varepsilon(2\ell+1)\tau}$ and
$e^{-iW\tau+i\varepsilon(2\ell-1)\tau}$ in Eq. (49) without expanding them. These are the complete transition
phases: each combines ordinary detuning with the kinetic-energy difference between its neighboring recoil
states. Alternatively solve Eq. (47) directly, with a converged ladder cutoff.

Before using a Wigner distribution, it is useful to state what changes when the initial electron is a wave
packet. For vacuum input, write the physical state in Eq. (45) as

\begin{equation}
|\Psi(t)\rangle=\sum_{\ell=0}^\infty\int dP\,\chi_\ell(P,t)|P,\ell\rangle,\quad \chi_\ell(P,t)=\phi(P+\ell q)A_\ell^{(0)}(P+\ell q,t).
\label{eq:83}
\end{equation}

A branch labelled $\ell$ is the component with exactly $\ell$ photons emitted into the initially empty
selected mode. Its final electron momentum $P$ is the initial momentum $p=P+\ell q$ minus $\ell q$. Here $t$
is physical time and $\tau=t/T_{0}$ its dimensionless counterpart; $C$ takes $\tau$ as its argument, while
$\chi$ includes the physical free phase and is written as a function of $t$. They describe the same instant.
The notation $\rho_{\ell m}^{(r)}=\langle\ell|\rho_{r}|m\rangle$ denotes a radiation density-matrix element;
$r$ labels radiation and $\ell,m$ its row and column photon numbers. Tracing over the appropriate factor gives

\begin{equation}
\begin{aligned}
\rho_{\ell m}^{(r)}(t)&=\int dP\,\chi_\ell(P,t)\chi_m^*(P,t),\\
\rho_e(P,P',t)&=\sum_\ell\chi_\ell(P,t)\chi_\ell^*(P',t),\\
P_\ell(t)&=\int dp\,|\phi(p)|^2|C_\ell^{(0)}(p,t/T_0)|^2.
\end{aligned}
\label{eq:84}
\end{equation}

These formulas separate the roles of the initial state. Photon-number probabilities for vacuum input average
the conditional probabilities over $f_{0}(p)=|\phi(p)|^{2}$, whereas optical off-diagonal elements depend on
overlap of the recoil-shifted electron packets. A coherent packet and an incoherent momentum mixture with the
same $f_{0}$ therefore give the same unconditional photon-number probabilities in this translation-invariant
vacuum-input model, but need not give the same optical coherence or electron spatial structure. This
equivalence does not extend automatically to a coherent optical seed, where different initial photon sectors
can interfere.

For illustration, a minimum-uncertainty Gaussian electron packet has

\begin{equation}
\phi(p)=\frac{1}{(2\pi\sigma_{p0}^2)^{1/4}}\exp\left[-\frac{(p-p_c)^2}{4\sigma_{p0}^2}-\frac{ipz_c}{\hbar}\right],\qquad\sigma_{z0}=\frac{\hbar}{2\sigma_{p0}}.
\label{eq:85}
\end{equation}

If the conditional physical amplitudes vary negligibly across the relevant shifted packet supports, the
overlap factor multiplying $A_{\ell}A_{m}^{\ast}$ is

\begin{equation}
\int dP\,\phi(P+\ell q)\phi^*(P+mq)=\exp\left[-\frac{(\ell-m)^2q^2}{8\sigma_{p0}^2}-\frac{i(\ell-m)qz_c}{\hbar}\right].
\label{eq:86}
\end{equation}

For $\sigma_{p0}\ll q$, different recoil orders leave nearly distinguishable electron states and optical
coherences are suppressed. For broader packets their overlap is larger, but the spread of detunings may then
spoil the assumption of constant amplitudes. The full integrals in Eq. (84), including the momentum-dependent
free phase, must be used whenever this approximation fails.

Initial uncertainty and interaction-generated fluctuations must consequently be kept distinct. The Gaussian
has an initial momentum variance $\sigma_{p0}^{2}$ and a corresponding position uncertainty before any FEL
emission. Recoil subsequently changes its momentum populations and correlations. For a phase-defined optical
input, initial electron coherence can also contribute immediately to the photon change:

\begin{equation}
\left.\frac{d\langle\hat n_r\rangle}{dt}\right|_{t=0}=2g\operatorname{Im}\left[\alpha_0^*\langle\hat T_q\rangle_0\right].
\label{eq:87}
\end{equation}

Explicitly, since $[\hat{a}^{\dagger},\hat{n}_{r}]=-\hat{a}^{\dagger}$ and $[\hat{a},\hat{n}_{r}]=\hat{a}$,
the Heisenberg equation gives

\begin{equation*}
\frac{d\langle\hat n_r\rangle}{dt}=ig\langle-\hat a^\dagger\hat T_q+\hat a\hat T_q^\dagger\rangle.
\end{equation*}

For the initial product state the expectation factors, with $\langle a\rangle=\alpha_{0}$ and $\langle
T_{q}^{\dagger}\rangle=\langle T_{q}\rangle^{\ast}$. Substitution yields Eq. (87). It vanishes for vacuum
input or a momentum-diagonal electron state. Thus the weak-transition number-state expressions Eq. (64) and
Eq. (65) cannot simply be reused for an arbitrary phase-localized electron and coherent seed by replacing $n$
with $|\alpha_{0}|^{2}$.

A second limitation is that the interaction itself changes the electron momentum distribution. Define

\begin{equation}
\sigma_p^2(t)=\langle p^2\rangle-\langle p\rangle^2,\qquad\Delta(p)=\frac{E(p-q)-E(p)}{\hbar}+\omega.
\label{eq:88}
\end{equation}

For vacuum input, the joint outcome distribution of initial momentum and exchanged photon number is
$|\phi(p)|^{2}|C_{\ell}^{(0)}(p,t/T_{0})|^{2}$. Since the outgoing momentum is $P=p-q\ell$, its variance
satisfies

\begin{equation}
\sigma_p^2(t)=\sigma_{p0}^2+q^2\Var(\ell)-2q\operatorname{Cov}(p,\ell).
\label{eq:89}
\end{equation}

The covariance is evaluated in that joint outcome distribution. If the exchange probabilities are independent
of the initial momentum, it vanishes and the recoil contribution adds to the initial variance. If detuning
selects parts of the packet, the covariance must be retained; a broadening formula based only on $q^{2}\Var
(\ell)$ is then incomplete.

Near a reference momentum $p_{r}$,

\begin{equation}
\Delta(p)\simeq\Delta(p_r)+\Delta'(p_r)(p-p_r),\qquad\sigma_\Delta(t)\simeq|\Delta'(p_r)|\sigma_p(t).
\label{eq:90}
\end{equation}

The two dispersions are physically distinct: recoil changes the neighboring ladder mismatch, while electron
momentum spread distributes the mismatch across the beam. Both can alter observable quantum signatures. The
initial width exists before radiation is exchanged, while the kinetic recoil correction changes the energy
spacing of successive exchange states. They can produce similar changes in a measured spectrum, but have
different origins and cannot be represented by the same parameter.

For an externally specified initial distribution one can average conditional probabilities,

\begin{equation}
\overline P_\ell(t)=\int dp\,f_0(p)P_\ell(t\mid p).
\label{eq:91}
\end{equation}

For vacuum input this is exactly the photon-number marginal in Eq. (84), for either a pure packet or a
momentum mixture with diagonal density $f_{0}$. Phase-sensitive observables require the density-matrix
information as well. This averaging does not imply that the outgoing electrons remain unchanged. Low-gain
response already produces perturbative energy modulation and distribution changes. The present calculation
describes one quantum electron, or an independent statistical ensemble of such single-electron passages in a
prescribed field. It does not include collective gain. A many-electron field evolving self-consistently with
electron correlations is a separate extension, and cannot be inferred from this single-electron average.

The Gaussian packet also makes the initial phase mixing explicit. Linearizing $W(p)$ about $p_{c}$, define
$W_{c}=W(p_{c})$ and $\sigma_{W}=|\partial_{p}W(p_{c})|\sigma_{p0}$. The weak-transition recoilless line shape
averaged over this packet is

\begin{equation}
\left\langle\mathcal S_0(W(p),\tau)\right\rangle_{f_0}=\int_0^\tau du\int_0^\tau dv\,\exp\left[iW_c(u-v)-\frac{\sigma_W^2(u-v)^2}{2}\right].
\label{eq:92}
\end{equation}

The quantity averaged here is the conditional finite-time spectral factor
$\mathcal{S}_{0}(W,\tau)=\tau^{2}\sinc ^{2}(W\tau/2)$ of Eq. (59), rather than a radiation-field amplitude
averaged over distinct electrons. For vacuum input, weak coupling and zero recoil, the selected-mode
spontaneous probability is $\mathfrak{g}^{2}\langle\mathcal{S}_{0}\rangle_{f_0}$ to leading order. Thus the
answer is indeed the spontaneous-emission line shape convolved with the entrance momentum density. For a fixed
initial photon number $n$, the same zero-recoil kernel multiplies $(n+1)\mathfrak{g}^{2}$ for emission and
$n\mathfrak{g}^{2}$ for absorption. It therefore also appears in stimulated transitions; the interpretation as
purely spontaneous applies specifically to vacuum input. Recoil separates the emission and absorption
detunings as established earlier.

This follows by performing the Gaussian momentum average of the two-time phase factor. The factor involving
$\sigma_{W}$ suppresses contributions separated by more than $1/\sigma_{W}$ in the dimensionless time
$\tau=t/T_{0}$; the corresponding physical separation is $T_{0}/\sigma_{W}$. With $\sigma_{W}=0$, the
conditional line shape is recovered. With recoil retained and nearly constant across the packet, the
elementary emission or absorption kernel uses the corresponding shifted central detuning. Thus initial
momentum uncertainty changes the finite-time spectrum as well as the visibility of individual recoil orders.

Changing variables to $s=u-v$ gives a convenient real integral for the numerical comparison,

\begin{equation}
\langle\mathcal S_0\rangle_{f_0}=2\int_0^\tau(\tau-s)\cos(W_c s)\,e^{-\sigma_W^2s^2/2}\,ds.
\label{eq:93}
\end{equation}

\begin{figure}[tbp]
\centering
\includegraphics[width=.82\linewidth]{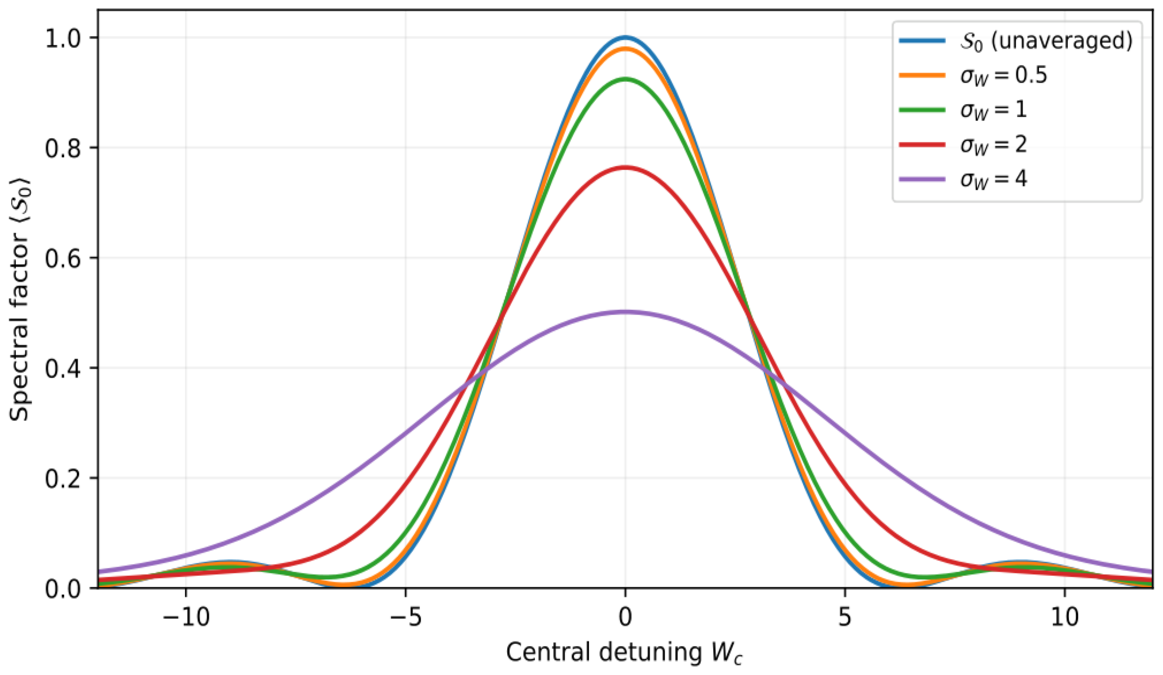}
\caption{Conditional and Gaussian-averaged zero-recoil line shapes at $\tau=1$. The solid reference is $\mathcal S_0(W_c,1)=\sinc^2(W_c/2)$; the other curves use Eq.~\eqref{eq:93} with dimensionless rms detuning widths $\sigma_W=0.5,1,2,4$. The ordinate is the spectral factor, without the coupling or photon-number multiplier.}
\label{fig:lineshapes}
\end{figure}

Figure~\ref{fig:lineshapes} compares this expression at $\tau=1$ with the unaveraged kernel for several
entrance widths. Momentum averaging lowers the central peak and smooths the zeros and side lobes; it broadens
the spectrum without increasing its integrated weight. Indeed, $\int
dW_{c}\langle\mathcal{S}_{0}\rangle=2\pi\tau$ for every normalized Gaussian convolution. A change of peak
height or spectral width in this calculation is consequently an entrance-distribution effect and does not by
itself identify quantum recoil.

The initial-state question has an operational consequence. Conventional beam measurements of energy spread,
emittance and bunch duration generally do not reconstruct the complete single-electron density operator. In
particular, the ensemble momentum spread need not equal the intrinsic momentum uncertainty of each electron
packet: it can include variation of packet centers from electron to electron. Choosing every electron as a
minimum-uncertainty Gaussian, or every electron as a sharp momentum state, is therefore a model assumption
unless the preparation or additional measurements support it. The uncertainty relation bounds compatible
variances but does not select a pure state or impose equality. Nor is a decomposition of a mixed density
operator into particular packets uniquely physical; all predictions depend on the density operator itself.

This incomplete specification does not make the recoil Hamiltonian or quantum corrections intrinsically
ambiguous. It makes some predictions conditional on information about the preparation. The distinction is
especially clear in the vacuum-input, translation-invariant problem already solved. For any normalized initial
electron density operator, its photon-number probabilities are

\begin{equation}
P_\ell(t)=\int dp\,\rho_{e0}(p,p)|C_\ell^{(0)}(p,t/T_0)|^2.
\label{eq:94}
\end{equation}

The momentum diagonal is understood as a probability density. Pure packets and mixed preparations with the
same diagonal give the same counts, even when their position coherence is different. The conditional
recoil-dependent line shape and its average are consequently meaningful without assuming a minimum-uncertainty
packet. This statement uses vacuum input, the specified interaction and unconditional photon counting; it is
not a state-independence theorem for all FEL observables.

Optical phase, electron bunching and coherently seeded response require more information. For example,

\begin{equation}
\langle\hat T_q\rangle_0=\int dp\,\rho_{e0}(p,p-q)
\label{eq:95}
\end{equation}

contains electron momentum coherence at separation $q$. It is not fixed by the measured momentum diagonal.
Randomizing the spatial centers of localized packets can leave that diagonal unchanged while suppressing Eq.
(95); Eq. (87) then gives a different phase-sensitive response. Our Gaussian and sharp-momentum examples
illustrate these distinct possibilities rather than competing claims about an unmeasured beam state. A fully
delocalized ideal momentum state must also not be confused with the measured finite duration of an ensemble
bunch.

A useful way to report this preparation uncertainty is to define the set $\mathcal{S}$ of positive, normalized
initial density operators compatible with the measured beam information. For an observable $O$, its predicted
range is

\begin{equation}
O_{\min}=\inf_{\rho_{e0}\in\mathcal S}O[\rho_{e0}],\qquad O_{\max}=\sup_{\rho_{e0}\in\mathcal S}O[\rho_{e0}].
\label{eq:96}
\end{equation}

A classical comparison uses a statistical ensemble constrained by the entrance measurements, rather than a
classical object representing an arbitrary pure quantum state. For the minimum-uncertainty Gaussian used here,
the initial Wigner function is positive and can be taken directly as a normalized classical phase-space
density with the same position and momentum marginals. The quantum prediction evolves it with the joint
Hamiltonian or the appropriate Wigner equation; the classical prediction evolves the same density with
Liouville transport and, for collective gain, the same Maxwell--beam feedback. Both calculations use the same
seed, cavity losses, mode selection and detector response. In a semiclassical photon-count baseline, a
conditional coherent intensity $\mu$ yields Poisson counts; its classical fluctuations must also be averaged.

For a non-Gaussian quantum preparation with a negative Wigner function there is no identical positive
classical phase-space density. One instead compares the measured observable with the family of positive
classical distributions compatible with the independently measured entrance data. Such a comparison is
operational, and does not assume that every quantum preparation admits a classical representation. The
particular baseline must be stated: otherwise a difference may be due to unequal entrance information rather
than the recoil dynamics being tested. A proposed recoil signature should therefore be compared with classical
predictions using compatible initial beam data and the same propagation and detection model. If the ranges
overlap within experimental uncertainty, that observable does not establish the proposed quantum correction
under the available characterization. One must either constrain the missing coherences or use an observable
insensitive to them. Recoil-dependent unconditional counts in Eq. (94) provide an example of the latter within
the model of one electron traversing the interaction region once with the selected radiation mode initially in
vacuum. A reliably measured $F<1$ is a different kind of evidence: it excludes a classical statistical mixture
of coherent radiation states, but does not by itself identify the initial electron state or establish that a
particular recoil mechanism caused it. Unknown preparation thus limits inference for specified observables; it
does not invalidate the quantum description.

A dimensionless dephasing measure is

\begin{equation}
\mathcal D(t)=\sigma_\Delta(t)T_{\rm coh},
\label{eq:97}
\end{equation}

where $T_{\mathrm{coh}}$ is the physical interaction or coherence time in seconds. With normalized frequencies
one equivalently uses $\sigma_{\Delta T_0}(\tau)(T_{\mathrm{coh}}/T_{0})$. Here $\sigma_{\Delta}$ is an
angular-frequency spread in $\mathrm{s}^{-1}$, so its product with a time in seconds is dimensionless. If an
energy-mismatch spread $\sigma_{\delta E}$ is used instead, the same measure is $\sigma_{\delta
E}T_{\mathrm{coh}}/\hbar$. For a static Gaussian mismatch distribution, $\langle e^{i\Delta
t}\rangle=e^{i\langle\Delta\rangle t-\sigma_\Delta^2 t^2/2}$, so $\mathcal{D}\ll1$ implies weak phase
dispersion over the coherence interval, whereas $\mathcal{D}\gtrsim1$ implies appreciable phase mixing.
Equation (92) derives this estimate explicitly in the present model. The width must be followed during the
interaction: a narrow entrance packet does not guarantee a narrow distribution after stimulated modulation or
spontaneous diffusion.

\section{Wigner phase space, modulation and bunching}

The Wigner representation provides a bridge between the recoil ladder and classical phase-space transport
\cite{ref40,ref41,ref42}. To avoid treating a dimensionless phase as a dimensional canonical coordinate, we
first use the pair $(z,p)$ with $[\hat{z},\hat{p}]=i\hbar$ and define

\begin{equation}
\mathcal W(z,p,t)=\frac1{2\pi\hbar}\int d\xi\,e^{-ip\xi/\hbar}\left\langle z+\frac\xi2\right|\hat\rho_e(t)\left|z-\frac\xi2\right\rangle.
\label{eq:98}
\end{equation}

For the vacuum-input packet already propagated in Section VI, its momentum-space form follows directly from
Eq. (84):

\begin{equation}
\mathcal W(z,p,t)=\frac1{2\pi\hbar}\sum_\ell\int d\kappa\,e^{i\kappa z/\hbar}\chi_\ell(p+\kappa/2,t)\chi_\ell^*(p-\kappa/2,t).
\label{eq:99}
\end{equation}

The variable $\kappa$ is a momentum difference. This is the Weyl representation of the electron density matrix
obtained from the Raman--Nath solution, not an independently postulated initial distribution. For the Gaussian
Eq. (85), its initial value is

\begin{equation}
\mathcal W_0(z,p)=\frac1{\pi\hbar}\exp\left[-\frac{(p-p_c)^2}{2\sigma_{p0}^2}-\frac{2\sigma_{p0}^2(z-z_c)^2}{\hbar^2}\right].
\label{eq:100}
\end{equation}

At later times Eq. (99) retains the recoil-branch weights, packet spreading and momentum coherences supplied
by the conditional amplitudes. A semiclassical closed Wigner equation, discussed below, is a further
approximation to this fully joint construction.

Its marginals are the position and momentum probability densities. It is normalized for a normalized electron
state but need not be positive everywhere; it is a quasiprobability. The phase $\theta=qz/\hbar$ can
subsequently be introduced with a corresponding canonical momentum rescaling.

For Weyl symbols $A$ and $B$, the star product and Moyal bracket are defined by \cite{ref41}

\begin{equation}
A\star B=A\exp\left[\frac{i\hbar}{2}\left(\overleftarrow\partial_z\overrightarrow\partial_p-\overleftarrow\partial_p\overrightarrow\partial_z\right)\right]B,
\label{eq:101}
\end{equation}

\begin{equation}
\{A,B\}_{\rm MB}=\frac{A\star B-B\star A}{i\hbar}.
\label{eq:102}
\end{equation}

The arrows state which factor is differentiated. The operator evolution
$i\hbar\dot{\rho}_{e}=[\hat{H},\rho_{e}]$ is the von Neumann equation. Its Weyl transform is the Wigner--Moyal
equation below. The finite-difference potential term derived subsequently is a contribution to that
transformed equation, rather than a separate operator von Neumann equation. The Moyal bracket approaches the
Poisson bracket as the higher-order terms become negligible. Applying the Weyl transform to
$\dot{\rho}_{e}=-i[\hat{H},\rho_{e}]/\hbar$ gives

\begin{equation}
\partial_t\mathcal W=\{H,\mathcal W\}_{\rm MB},\qquad H(z,p,t)=E(p)+V(z,t).
\label{eq:103}
\end{equation}

This exact equation assumes a specified scalar electron Hamiltonian. In an entangled electron--field problem
the reduced electron equation need not close on $\mathcal{W}$ alone. Its coupling contains field-conditioned
electron correlations unless a semiclassical or mean-field reduction is made \cite{ref25,ref33}.

The potential coefficient formerly left schematic can be derived explicitly. In a semiclassical field
approximation replace $\hat{a}$ by a complex amplitude $\alpha(t)$. The interaction in Eq. (19) becomes

\begin{equation}
V(z,t)=V_-(t)e^{-iqz/\hbar}+V_-^*(t)e^{iqz/\hbar},\qquad V_-(t)=\hbar g\alpha^*(t).
\label{eq:104}
\end{equation}

All quantities are dimensional here. The star-product translation identities are

\begin{equation}
\begin{aligned}
e^{-iqz/\hbar}\star\mathcal W&=e^{-iqz/\hbar}\mathcal W(z,p+q/2,t),\\
\mathcal W\star e^{-iqz/\hbar}&=e^{-iqz/\hbar}\mathcal W(z,p-q/2,t).
\end{aligned}
\label{eq:105}
\end{equation}

For the conjugate harmonic the two shifts are interchanged. Substitution into Eq. (102) gives the exact
potential contribution

\begin{equation}
\begin{aligned}
\left.\partial_t\mathcal W\right|_V&=\frac{V_-e^{-iqz/\hbar}-V_-^*e^{iqz/\hbar}}{i\hbar}[\mathcal W(z,p+q/2,t)-\mathcal W(z,p-q/2,t)]\\
&=2g\operatorname{Im}[\alpha^*(t)e^{-iqz/\hbar}][\mathcal W(z,p+q/2,t)-\mathcal W(z,p-q/2,t)].
\end{aligned}
\label{eq:106}
\end{equation}

Thus the coefficient, sign and normalization follow from the same Hamiltonian. Quantum recoil appears as a
finite momentum translation. The shift is $q/2$ in the Wigner equation because the density matrix combines two
momentum arguments; the elementary electron transition still changes momentum by $q$.

For a quadratic kinetic approximation $E(p)\simeq E_{0}+v_{0}(p-p_{0})+(p-p_{0})^{2}/(2m_{\parallel})$, the
free contribution is exactly $-[v_{0}+(p-p_{0})/m_{\parallel}]\partial_{z}\mathcal{W}$. For the full
relativistic dispersion the kinetic star commutator must also be retained. Its derivative expansion begins

\begin{equation}
\left.\partial_t\mathcal W\right|_E=-E'(p)\partial_z\mathcal W+\frac{\hbar^2}{24}E'''(p)\partial_z^3\mathcal W+\cdots.
\label{eq:107}
\end{equation}

It would be inconsistent to describe the potential term as exact while silently replacing a non-quadratic
kinetic term by purely classical drift.

Expanding the finite difference in Eq. (106),

\begin{equation}
\mathcal W(p+q/2)-\mathcal W(p-q/2)=q\partial_p\mathcal W+\frac{q^3}{24}\partial_p^3\mathcal W+\frac{q^5}{1920}\partial_p^5\mathcal W+\cdots.
\label{eq:108}
\end{equation}

Since $\partial_{z}V=q[V_{-}e^{-iqz/\hbar}-V_{-}^{\ast}e^{iqz/\hbar}]/(i\hbar)$, the leading potential term is
$\partial_{z}V\partial_{p}\mathcal{W}$. Together with the leading kinetic term it gives

\begin{equation}
\partial_t\mathcal W+E'(p)\partial_z\mathcal W-\partial_zV\,\partial_p\mathcal W=0,
\label{eq:109}
\end{equation}

the classical Liouville/Vlasov transport equation for the specified Hamiltonian.

A closely related low-gain phase-space analysis is given by Carmesin, Kling, Giese, Sauerbrey and Schleich
\cite{ref43}. They compare quantum Wigner evolution with classical trajectories in the same periodic potential
and show that populating many momentum levels is not sufficient for classical correspondence. In their
notation $\alpha_{W}=U_{0}/(2\hbar\omega_{r})$ measures potential height relative to recoil energy; it is
distinct from the oscillator parameter $\alpha_{n}$ of Section XII. A large $\alpha_{W}$ must be accompanied
by a distribution smooth on the recoil momentum scale. Their derivative expansion is therefore consistent with
the finite-difference criterion derived below. They also show that phase-space quantum structure can be
suppressed in averaged observables such as gain. This complements our comparison between RN ladder amplitudes,
reduced electron distributions and radiation statistics: agreement of a gain curve does not imply agreement of
the full quantum state. The importance of this work is therefore both conceptual and experimental: it
identifies which phase-space features are removed by beam averaging and motivates measurements beyond an
integrated first moment. It is a phase-space analysis, not itself a demonstration of a few-photon cavity
experiment. A complementary three-dimensional discrete-Wigner treatment by Piovella and collaborators includes
longitudinal momentum quantization and transverse geometry \cite{ref44}; it addresses a broader beam-evolution
problem than the low-gain single-electron comparison considered here.

Their phase-uniform initial distribution, $\mathcal{W}(\theta,\wp ;0)=\rho(\wp )/(2\pi)$, represents a
statistical ensemble; position and momentum widths need not be tied as in a pure minimum-uncertainty packet.
Our RN packet calculations retain a specified quantum input and its recoil coherences, whereas their Wigner
treatment emphasizes ensemble averaging and uses a semiclassical radiation amplitude. It is consequently a
useful comparison for electron phase-space dynamics, not a calculation of the fully quantized cavity photon
distribution. The two approaches share the recoil ladder but answer different observable questions.

The higher odd derivatives give quantum corrections. Replacing the potential difference by its first
derivative requires smoothness on the recoil scale,

\begin{equation}
R_q=\frac{q}{\Delta p_{\rm char}}\ll1.
\label{eq:110}
\end{equation}

Here $\Delta p_{\mathrm{char}}$ is the local momentum scale over which the Wigner function varies appreciably
in the region contributing to the observable. It is not automatically the rms beam spread: interference
fringes or narrow substructure can have much smaller scales. Quantitatively, where derivative ratios are
meaningful, $(\Delta p_{\mathrm{char}})^{-2}\sim|\partial_{p}^{3}\mathcal{W}/\partial_{p}\mathcal{W}|$ gives
the leading correction relative to the first derivative. The leading omitted potential term is then of
relative order $q^{2}/(24\Delta p_{\mathrm{char}}^{2})$.

For nonquadratic free dispersion a spatial smoothness condition must also be checked. If $L_{\mathrm{char}}$
is the spatial scale of variation of $\mathcal{W}$, the leading kinetic correction relative to classical drift
is of order

\begin{equation}
R_z\sim\frac{\hbar^2}{24L_{\rm char}^2}\left|\frac{E^{(3)}(p)}{E'(p)}\right|\ll1.
\label{eq:111}
\end{equation}

This estimate applies where the leading drift is nonzero; otherwise the kinetic terms must be compared
directly. Higher derivatives require analogous checks. For a strictly quadratic $E$, all higher kinetic
corrections vanish, so no additional spatial restriction follows from that part of the Moyal expansion. This
criterion concerns spatial structure of the Wigner function relative to the dispersion scale, and does not
universally require a packet longer than the optical grating period. This limit is conceptually distinct from
the phase expansion in Eq. (51). The latter controls a finite-time approximation to the Raman--Nath hopping
phases; the former controls a phase-space derivative expansion. The neighboring-state energy differences
already derived in Eq. (35) and Eq. (50) supply their common recoil scale, so they need not be repeated here.

The first momentum moments and bunching are

\begin{equation}
\langle p\rangle=\int dz\,dp\,p\mathcal W,\quad \sigma_p^2=\int dz\,dp\,(p-\langle p\rangle)^2\mathcal W,\quad b_-=\int dz\,dp\,e^{-iqz/\hbar}\mathcal W.
\label{eq:112}
\end{equation}

Introducing the Colson current means that the field equation describes a many-electron beam response. It is
not a single-electron source equation. The electron distribution supplies bunching, and the common field
changes that distribution in return. With a specified FEL normalization and field phase convention one may
write schematically $da/d\tau=-j_{c}b_{-}$, where $j_{c}$ is the Colson current. Numerical coefficients and
field-phase conventions must be fixed by the adopted normalization. Single-electron quantum evolution enters
the collective problem through the beam distribution or density matrix and its correlations.

Increasing momentum spread can make individual recoil steps less resolvable while increasing the distribution
of detunings and phase mixing. The three distinct controls are recoil resolution $R_{q}$, accumulated local
phase $R_{\phi}\sim|\varepsilon|\ell\tau$, and dephasing $\mathcal{D}$ from Eq. (97). A large spread can
suppress visible quantum structure and simultaneously reduce classical FEL gain. Formal substitution
$\varepsilon\to0$ does not describe all these effects.

To see explicitly which electron information is required for bunching, consider a pure electron superposition
in a periodic length $L$,

\begin{equation}
\psi_e(z,t)=\frac1{\sqrt L}\sum_\ell c_\ell(t)e^{i(p_0-\ell q)z/\hbar}.
\label{eq:113}
\end{equation}

Its Wigner function is

\begin{equation}
\mathcal W=\frac1L\sum_{\ell,j}c_\ell c_j^*e^{-i(\ell-j)qz/\hbar}\delta\left[p-p_0+\frac{(\ell+j)q}{2}\right].
\label{eq:114}
\end{equation}

Diagonal terms give momentum populations; off-diagonal terms give spatial modulation. Integration over an
integer number of periods yields

\begin{equation}
b_+=\langle e^{iqz/\hbar}\rangle=\sum_j c_{j+1}c_j^*,\qquad b_-=b_+^*.
\label{eq:115}
\end{equation}

For a mixed electron state the products are replaced by density-matrix elements. By contrast, tracing over
photons in the sharp-momentum, sharp-photon-number joint state Eq. (43) gives

\begin{equation}
\rho_e=\sum_\ell |C_\ell|^2|p_0-\ell q\rangle\langle p_0-\ell q|.
\label{eq:116}
\end{equation}

This particular preparation has no unconditional spatial bunching. For the initial packet, however, Eq. (84)
retains coherences within each recoil branch, and the corresponding bunching is

\begin{equation}
b_-(t)=\sum_\ell\int dP\,\chi_\ell(P,t)\chi_\ell^*(P-q,t).
\label{eq:117}
\end{equation}

For the initial Gaussian it is $b_{-}(0)=e^{-q^2/(8\sigma_{p0}^2)}e^{-iqz_c/\hbar}$. This may be nonzero
before the interaction because the packet is spatially localized; it must be distinguished from FEL-induced
modulation. The subsequent variation follows from the same Raman--Nath branch amplitudes used in Eq. (99). An
ensemble with uniformly distributed centers $z_{c}$ has vanishing initial mean bunching even when each
individual packet is localized. A phase-defined field, wave packets, conditional measurements or a
self-consistent many-electron preparation can supply electron coherences. Joint electron--field correlations
can remain important even when the reduced electron bunching vanishes. Electron wave-function modulation is
therefore preparation-dependent, not an automatic interpretation of every recoil population.

\section{Photon-exchange bound, field coupling and the pendulum}

The local mismatch in Eq. (50) gives a transparent recoil bound. Recoil is negligible relative to a nonzero
detuning only when

\begin{equation}
|\varepsilon(2\ell\pm1)|\ll|W|,\qquad\ell\ll\ell_r\sim\frac{|W|}{|\varepsilon|}.
\label{eq:118}
\end{equation}

The estimate suppresses an order-unity factor of two. It measures how many recoil steps can be explored before
the local resonance shift is comparable to the original detuning. Close to $W=0$ this comparison fails; the
reference scale must instead be the finite-time width, gain bandwidth or coherence frequency. It must not be
interpreted as a singular physical recoil constraint at exact resonance.

The dimensional neighboring photon-state matrix element is $\hbar g\sqrt{n+1}:\sqrt{n+1}$ follows from
$\hat{a}^{\dagger}|n\rangle=$ $\sqrt{n+1}|n+1\rangle$, whereas absorption carries $\sqrt n$. It is not $n+1$
without the square root. In the dimensionless ladder its coupling is $\mathfrak{g}\sqrt{n+1}$. At a
photon/recoil scale $n_{r}\sim|W|/|\varepsilon|$, introduce

\begin{equation}
\Lambda=\frac{\G\sqrt{n_r}}{|W|}\sim\frac{\G}{\sqrt{|W\varepsilon|}}.
\label{eq:119}
\end{equation}

For the explicit equal-frequency mode normalization of Eq. (10), put $\Delta=W/T_{0}$ and
$\omega_{q}=\varepsilon/T_{0}=$ $q^{2}/(2m\hbar)$ in the undressed nonrelativistic BR model. The pump
occupation is $\overline{\nu_u}$. Then

\begin{equation}
\Lambda\sim\frac{g}{\sqrt{|\Delta|\omega_q}}=\frac{\alpha_{\rm fs}c\bar\lambda_C\lambda'|\xi|\sqrt{\overline{\nu_u}}}{\sqrt{V'_uV'_r}\sqrt{|\Delta|\omega_q}}.
\label{eq:120}
\end{equation}

For an optical pump $q=2\hbar\omega^{\prime}/c$, this gives

\begin{equation}
\frac{n_r}{n_{\rm mix}}\sim\Lambda^2=\frac{\alpha_{\rm fs}^2c\bar\lambda_C(\lambda')^4|\xi|^2\overline{\nu_u}}{8\pi^2V'_u V'_r|\Delta|}.
\label{eq:121}
\end{equation}

Here $\omega_{q}=8\pi^{2}c\bar{\lambda}_{C}/(\lambda^{\prime})^{2}$ for the optical-pump recoil
$q=4\pi\hbar/\lambda^{\prime}$. The ordinary Compton-wavelength form follows by replacing $\bar{\lambda}_{C}$
with $\lambda_{C}/(2\pi)$. These expressions expose the electromagnetic coupling but also its dependence on
pump occupation, overlap and radiation-mode volume. The factor $\overline{\nu_u}/V_{u}^{\prime}$ is the
physical pump density, so an arbitrary normalization-volume choice does not alter the result. A laboratory
relativistic or transversely dressed model must use its own kinetic curvature and coupling, rather than
transplant this BR prefactor.

A numerical volume estimate requires a mode geometry. For a traveling Gaussian mode whose transverse field
profile is $e^{-r^2/{w^{\prime}}^2}$, approximately constant over a length $L_{j}^{\prime}$, the
intensity-normalized effective volume is

\begin{equation}
V'_j=\int|u_j(\boldsymbol r')|^2d^3r'\simeq\frac{\pi {w'_j}^2}{2}L'_j,\qquad\max|u_j|=1.
\label{eq:122}
\end{equation}

For example, $w_{j}^{\prime}=10\mu$m and $L_{j}^{\prime}=1$ mm give $V_{j}^{\prime}\simeq1.57\times10^{-13}$
$\mathrm{m}^3$. With equal such volumes, $\lambda^{\prime}=10$ nm and $|\xi|=1$, the illustrative
single-quantum coupling is $|g_{u}|\simeq5.38\times10^{-2}$ $\mathrm{s}^{-1}$. This is a normalization
example, not a reconstruction of the Kling--Giese design. A standing-wave mode adds its longitudinal intensity
factor; a focused traveling mode requires the integral over its changing waist. The actual radiation-mode
volume is determined by cavity or wave-packet normalization, not automatically by the electron interaction
length. For a box-normalized traveling pump, changing $V_{u}^{\prime}$ and its assigned occupation together
leaves $\overline{\nu_u}/V_{u}^{\prime}$ and the physical drive unchanged. Consequently no unique physical
volume, or numerical $\Lambda$, follows from wavelength and interaction length alone: waist, mode profile,
cavity length and pump density must also be specified.

Thus $\Lambda\ll1$ indicates that neighboring-state mixing remains weak near the estimated recoil boundary.
This is a single-electron indicator, not a universal definition of FEL small signal.

A complementary strong-mixing scale follows from $\mathfrak{g}\sqrt{n_{\rm mix}}\sim|W|$:

\begin{equation}
n_{\rm mix}\sim\frac{W^2}{\G^2},\qquad\frac{n_r}{n_{\rm mix}}\sim\Lambda^2,\qquad n\ll\min\left(\frac{|W|}{|\varepsilon|},\frac{W^2}{\G^2}\right).
\label{eq:123}
\end{equation}

Since $n_{r}/n_{\mathrm{mix}}\sim\Lambda^{2},\Lambda\ll1$ means $n_{r}\ll n_{\mathrm{mix}}$: as the occupation
or accessible exchange scale increases, the recoil mismatch reaches the detuning scale while the matrix
element is still too small for strong mixing. Conversely, $\Lambda\gg1$ means $n_{\mathrm{mix}}\ll n_{r}$: the
coupling reaches the detuning scale before the estimated recoil boundary is reached. The statement orders two
crossover scales; it does not assert that the actual dynamics necessarily populates either scale. In
particular, photon occupation and net exchange order need not coincide for a strongly seeded field. These are
estimates for the adopted detuned ladder. They do not replace the time-dependent perturbative condition Eq.
(51) or establish classical trapping by themselves.

The classical single-electron FEL equations can be expressed, after fixing a pendulum normalization and
optical phase, as

\begin{equation}
\dot\zeta=\nu,\qquad\dot\nu=-|a|\sin(\zeta+\phi),\qquad H_p=\frac{\nu^2}{2}-|a|\cos(\zeta+\phi).
\label{eq:124}
\end{equation}

Here the dot is with respect to the normalized pendulum time. The cosine Hamiltonian gives the displayed sine
force with the indicated sign. A sine Hamiltonian would require the corresponding phase shift. The
small-amplitude bounce frequency obeys

\begin{equation}
\omega_B\sim\sqrt{|a|}.
\label{eq:125}
\end{equation}

In a coherent-field semiclassical identification the potential amplitude is proportional to $\mathfrak{g}\sqrt
n$, with constants fixed by the momentum and time normalization. Hence $\omega_{B}^{2}$, not $\omega_{B}$,
scales with that potential amplitude. A trapping comparison must use the free kinetic scale appropriate to the
orbit, generally involving the detuning momentum and kinetic curvature. The quantum level-mixing boundary in
Eq. (123) should therefore not be equated directly with a universal classical trapping threshold.

The comparison nevertheless clarifies the two ways the elementary regime can fail: accumulated recoil changes
the local resonance, or the field produces substantial nonlinear motion. Their relationship requires the full
normalization connecting $W,\varepsilon,\mathfrak{g},\nu$ and $a$.

\section{Canonical interaction picture and perturbative structure}

The classical interaction-picture construction discussed in Ref. \cite{ref33} reinforces the phase-space
interpretation. Let $H=H_{0}+V$ and define the Liouville generator by
$\mathcal{L}_{H}\rho=\{H,\rho\}_{\mathrm{PB}}$. Choosing a manageable reference flow $U_{0}$ gives

\begin{equation}
\rho_I=U_0^{-1}\rho,\qquad\partial_t\rho_I=\mathcal L_{1I}(t)\rho_I,\qquad\mathcal L_{1I}=U_0^{-1}\mathcal L_1U_0.
\label{eq:126}
\end{equation}

A time-ordered Neumann--Dyson expansion follows. This is the classical counterpart of removing the manageable
detuning evolution before expanding residual recoil phases. It explains why a local mismatch organization is
more useful than expanding the absolute diagonal quantity $\varepsilon\ell^{2}$.

The Wigner equation similarly separates classical transport and quantum corrections,

\begin{equation}
\partial_t\mathcal W=(\mathcal L_{\rm cl}+\mathcal L_q)\mathcal W,
\label{eq:127}
\end{equation}

where $\mathcal{L}_{q}$ contains the higher derivatives associated with the Moyal bracket. For a quadratic
free Hamiltonian and a spatial potential, those terms are the higher odd momentum derivatives in Eq. (108).
For a nonquadratic dispersion the kinetic corrections in Eq. (107) must also be included.

A finite Dyson or Neumann polynomial need not remain an exact unitary propagator or an exact canonical-flow
operator, even if the underlying dynamics has that structure. This does not mean that every finite-order
density expansion necessarily violates its integral normalization: the property depends on the generator and
truncation. A Magnus organization, with its exponent in the relevant Lie algebra, preserves the operator
structure more directly. In the quantum problem its exponent is anti-Hermitian; in a suitable classical
Hamiltonian-flow formulation its generator retains the canonical structure.

For the Raman--Nath probability expansion, order-by-order normalization requires

\begin{equation}
\sum_\ell P_\ell^{(0)}=1,\qquad\sum_\ell P_\ell^{(1)}=0,\qquad\sum_\ell P_\ell^{(2)}=0.
\label{eq:128}
\end{equation}

These conditions are compatible with Eq. (74) when the amplitude hierarchy is generated consistently. A finite
truncation need not be positive or uniformly accurate outside the controlled domain. A structure-preserving
treatment is particularly useful before iterating a perturbative channel through many cavity passages.

\section{From single-electron dynamics to beam response}

For independent incoming electrons with a normalized initial phase-space distribution $f_{0}(p,\theta)$, a
linear spectral response can be averaged as

\begin{equation}
\overline{\Gamma}_\pm(\Delta)=N_e\int dp\,d\theta\, f_0(p,\theta)\Gamma_\pm(p,\theta;\Delta).
\label{eq:129}
\end{equation}

This is useful for initial response and classical correspondence. It is not a claim that electrons remain
unmodified in low gain. The outgoing distribution acquires perturbative changes at the order at which energy
modulation and bunching are calculated.

Even a linearized high-gain calculation must retain the self-consistent field--distribution coupling. A
fixed-distribution convolution supplies an elementary beam-averaged kernel; sustained collective amplification
requires a kinetic closure or a many-electron quantum hierarchy. The relevant dynamical information includes
the field amplitude and photon number, electron bunching, mean momentum, momentum variance, and correlations
that are not determined by those moments alone.

A quantum treatment averages incoming electron density operators and transition channels. Electron shot noise,
vacuum emission and cavity reservoir noise must be included consistently, once each. An extra phenomenological
spontaneous seed can double count feeding already present in a quantum interaction map. The independent
classical shot-noise baseline is discussed in Appendix B; it remains essential when deciding whether a
measured fluctuation is a quantum signature.

\section{Quantum cavity evolution of the FEL oscillator}

A resonator stores the field while fresh electron bunches repeatedly interact with it. Its stationary state
depends on pumping, loss, saturation and mode overlap; it is not obtained by multiplying a single-electron
recoil correction by the number of passages.

The cavity research initiated by Walther provides a particularly relevant experimental precedent. Meschede,
Walther and Müller demonstrated photon exchange between individual Rydberg atoms and a superconducting
microwave cavity in the one-atom maser \cite{ref45}. Rempe, Walther and Klein subsequently observed quantum
collapse and revival, with velocity selection controlling the interaction time \cite{ref46}. These experiments
and the cavity-QED review by Walther and collaborators \cite{ref47} establish the value of preparing the
incident particle, controlling its transit, storing a small optical or microwave occupation, and inferring
field statistics from particle dynamics. The Jaynes--Cummings description \cite{ref48} supplies the elementary
atom--field model: occupation-dependent exchange frequencies can dephase and later rephase. Its formal
relation to an isolated electron recoil pair is useful, but the atomic internal transition and the
free-electron momentum ladder have different preparation and coupling requirements. Collapse and revival are
therefore a possible target only after coherence, interaction-time uniformity and cavity lifetime have been
established; they are not implied by a stationary FEL photon distribution.

Let $\rho_{m}$ be the intracavity field density operator before bunch $m$, and $\rho_{e,in}^{(m)}$ the state
of the fresh incoming electrons. The single-pass field channel is

\begin{equation}
\mathcal E_m(\rho_m)=\operatorname{Tr}_e\left[U_m(\rho_m\otimes\rho_{e,\rm in}^{(m)})U_m^\dagger\right].
\label{eq:130}
\end{equation}

The outgoing electrons are traced out because they leave the cavity. Their recoil and correlations with the
light affect the field channel, but their outgoing state is not reused as the next incoming bunch. For
independently prepared incoming bunches the exact channel is completely positive and trace preserving.

Including the cavity phase, losses and pulse timing gives

\begin{equation}
\rho_{m+1}=\mathcal L_{\eta,\phi_c,\delta L}[\mathcal E_m(\rho_m)].
\label{eq:131}
\end{equation}

The loss channel includes the appropriate environmental modes. For a single port with fractional power loss
$0\le\eta\le1$ per passage through that loss element, its operator representation is

\begin{equation}
\hat a_{\rm after}=\sqrt{1-\eta}\,\hat a_{\rm before}+\sqrt\eta\,\hat v,
\label{eq:132}
\end{equation}

where $\hat{v}$ is an independent reservoir mode. Vacuum noise is consequently an operator contribution. An
equivalent stochastic amplitude formulation must state its phase-space representation and noise prescription.

A selected single-mode model is an appropriate reduction when timing, spatial overlap and spectral selection
support it. Finite bunch duration, slippage, radiation lethargy and longitudinal mode competition can instead
require an envelope description. The long-bunch, low-gain pulse-propagation review in Ref. \cite{ref49}
provides the relevant envelope treatment. The evolution-operator and time-ordering methods of Ref.
\cite{ref36} provide its mathematical context, while Ref. \cite{ref50} discusses high-gain oscillator pulse
propagation and saturation. These extensions do not invalidate the selected-mode reduction when its
assumptions hold. In a classical envelope approximation,

\begin{equation}
A_{m+1}(\zeta)=\sqrt{1-\eta}\,e^{i\phi_c}\mathcal T_{\delta L}\{\mathcal U_{\rm FEL}[A_m,f_{e,\rm in}^{(m)}]\}(\zeta),\qquad\mathcal T_{\delta L}=e^{-\delta\zeta\partial_\zeta}.
\label{eq:133}
\end{equation}

Here $\eta$ is the total fractional intracavity power lost in one round trip, including output coupling and
internal losses, with $0\le\eta<1$ for an operating cavity. It is dimensionless and differs from an
amplitude-loss fraction: the amplitude transmission is $\sqrt{1-\eta}$. The symbol $\delta\zeta$ is the
dimensionless relative pulse displacement produced by the cavity mismatch $\delta L$, with its conversion
determined by the chosen longitudinal coordinate. Discrete-electron noise can enter a stochastic classical
treatment; a quantum treatment derives its fluctuations from the channel rather than appending an unspecified
random field.

An equilibrium pulse satisfies a functional fixed-point condition, up to an overall phase. If $G(I)$ is the
single-pass fractional intensity gain, its scalar intensity projection is

\begin{equation}
(1-\eta)[1+G(I_{\rm eq})]=1,\qquad G(I_{\rm eq})=\frac{\eta}{1-\eta}\simeq\eta.
\label{eq:134}
\end{equation}

The final approximation applies to small loss. This determines only a mean-intensity condition; photon
statistics require the stationary density operator.

\section{Comparison with Kling and Giese}

\subsection*{An innovative route toward dedicated quantum-FEL experiments}

Kling and Giese \cite{ref24} propose an interesting and innovative quantum FEL oscillator: a
counter-propagating optical undulator produces well-separated recoil levels, while successive fresh electron
bunches amplify radiation stored in a resonator. The short interaction region reduces the accumulated effects
of space charge and unwanted spontaneous emission compared with a long single-pass device. Their proposal
gives a constructive perspective for dedicated experiments on recoil-selective electron--photon exchange and
radiation statistics. The following discussion identifies the conditions needed to develop that perspective;
it does not imply that the proposed quantum regime is impossible.

The resonant pair in the BR frame is $|p,n\rangle\leftrightarrow |p-q,n+1\rangle$, with $p\simeq q/2$. For
recoil momentum $q=2\hbar k$, the recoil frequency and two-state selectivity parameter are

\begin{equation}
\omega_r=\frac{q^2}{2m\hbar},\qquad\alpha_n=\frac{g\sqrt{n+1}}{\omega_r}\ll1.
\label{eq:135}
\end{equation}

Here $g$ and $\omega_{r}$ are BR-frame angular frequencies and $T$ below is the interaction duration in that
frame. The optical-undulator pump is represented by a prescribed classical amplitude, consistent with the
undepleted, highly occupied coherent-pump limit discussed in Section VI. It must be distinguished from the
stored radiation, whose stationary state need not be coherent. A photon-number-state pump represents a
different preparation; artificial recoil-free antibunching from that choice is excluded by coherent-pump
averaging \cite{ref39}.

\subsection*{What the stationary photon distribution includes}

The two-state transition probability is

\begin{equation}
\begin{aligned}
\Omega_n(p)&=\sqrt{g^2(n+1)+\omega_r^2(p/q-1/2)^2},\\
P_{\rm em}(n,p)&=g^2(n+1)T^2\sinc^2[\Omega_n(p)T].
\end{aligned}
\label{eq:136}
\end{equation}

The $n+1$ factor includes the vacuum contribution and stimulated enhancement by the stored photons. The
$\sinc$ factor expresses finite-time resonance acceptance and coherent exchange; it does not mean that only
spontaneous radiation is stored. Reabsorption within the resonant pair is part of the same unitary evolution.

Under their independent-event pumping and cavity-damping assumptions, the stationary probabilities are

\begin{equation}
\begin{aligned}
P_n&=P_0\prod_{j=1}^n\left[\vartheta^2\int dp\,\rho_e(p)\sinc^2[\Omega_{j-1}(p)T]\right],\\
\vartheta&=gT\sqrt{N_a},\qquad N_a=\frac{N_eQ}{\omega_L\tau_{\rm inj}}.
\end{aligned}
\label{eq:137}
\end{equation}

$N_{e}$ denotes electrons per bunch, $Q$ the cavity quality factor and $\rho_{e}(p)$ the normalized entrance
momentum density. Thus entrance energy averaging is already present. The result is a stationary distribution
after repeated pumping and loss, not the distribution immediately after one round trip. In the birth--death
balance, the factor $j$ in emission from $j-1$ photons cancels the factor $j$ in cavity loss. Its absence from
the displayed product does not remove stimulated emission. Their supplementary derivation \cite{ref51} traces
out each outgoing electron; additional initial electron coherences or collective bunch correlations require
the appropriate joint channel.

The diagnostic is $F=\Var (n)/\langle n\rangle$. Near threshold, their resonant small-signal estimate is
$F_{q}\simeq1/\delta$, where $\delta=(G-\omega_{L}\tau_{\mathrm{inj}}/Q)/G$ and $G=N_{e}(gT)^{2}$. It is
therefore super-Poissonian for $0<\delta\ll1$, although narrower than their classical small-recoil comparison.
Sub-Poissonian regions, $F<1$, occur in suitable nonlinear operating regimes of their model. Reduced excess
fluctuations and nonclassical counting statistics should consequently be distinguished.

\subsection*{Reported beam parameters and the available acceptance}

Tables 1--2 of the proposal provide the quantitative starting point \cite{ref24}:

\begin{center}
\begin{tabular}{llll}
$\gamma_0$ & $51.8$ & $L$ & $1.14\,\mathrm{mm}$\\
$\lambda_W$ & $1.064\,\mu\mathrm m$ & $\lambda_L$ & $1$~\AA\\
$n_e$ & $6.39\times10^{22}\,\mathrm{m}^{-3}$ & $I_p$ & $8.9\,\mathrm A$\\
$\sigma_e$ & $0.68\,\mu\mathrm m$ & $\tau_e$ & $1.2\,\mathrm{ps}$\\
$Q_b$ & $11\,\mathrm{pC}$ & $\epsilon_n$ & $0.0082\,\mathrm{mm\,mrad}$\\
$\sigma_{\epsilon,0}$ & $7.5\times10^{-5}$ & $G$ & $0.1$
\end{tabular}
\end{center}

The quoted spread is $0.075$ per mille, not $0.075$ percent. Numerical width estimates below interpret it as
rms and interpret $\sigma_{e}$ as a one-plane rms waist size. The reported bunch-duration scale gives
$\ell_{b}\simeq$ $c\tau_{e}=0.360$ mm, about one-third of $L$; its precise profile convention must be retained
in a detailed overlap calculation. The optical pulse duration $7.6$ ps corresponds to $2.28$ mm, approximately
$2L$ for a counterpropagating pulse, and the Rayleigh length is $4.4$ mm. The article's slippage estimate
$(L/\lambda_{W})\lambda_{L}\simeq0.107\mu$m is much smaller than the bunch length.

The current and density are consistent with the transverse Gaussian convention

\begin{equation}
n_{e,0}=\frac{I_p}{2\pi e\beta c\sigma_x\sigma_y},\qquad I_p=\frac{Q_b}{\sqrt{2\pi}\sigma_t}\quad\text{for a Gaussian time profile.}
\label{eq:138}
\end{equation}

The second relation is a profile-specific identity, not an identification of the quoted $\tau_{e}$ with
$\sigma_{t}$. The reported density, radius and current give $I_{p}\simeq8.9$ A without assuming that
identification.

Writing $N_{W}=L/\lambda_{W}\simeq1071$ for the laboratory optical period count, their finite-time
electron-energy acceptance is

\begin{equation}
\Gamma=\frac{\lambda_W}{4\pi L}\simeq7.45\times10^{-5},\qquad\mu_{\epsilon,W}=4N_W\sigma_{\epsilon,0}\simeq0.321,\qquad\frac{\sigma_{\epsilon,0}}{\Gamma}\simeq1.
\label{eq:139}
\end{equation}

For comparison with the conventional magnetic-undulator detuning kernel, the counterpropagating wave has
effective period count $N_{\mathrm{eff}}\simeq2N_{W}$; hence $\mu_{\epsilon,eff}\simeq0.643$ and the
accumulated rms detuning phase is $2\sigma_{\epsilon,0}/\Gamma\simeq2.0$. These conventions must not be
interchanged. At $1$~\AA{} the photon energy is $12.40$ keV and

\begin{equation}
r=\frac{\hbar\omega_L}{\gamma_0mc^2}\simeq4.68\times10^{-4},\qquad\frac{\sigma_{\epsilon,0}}{r}\simeq0.16.
\label{eq:140}
\end{equation}

Entrance spread alone therefore permits recoil separation, while already occupying a substantial part of the
narrower resonance acceptance. A Gaussian, weak-field, on-resonance average of $\sinc
^{2}[(\delta\gamma/\gamma_{0})/\Gamma]$ gives about $0.76$ of the monoenergetic response. This is a diagnostic
estimate, not a recalculated cavity threshold.

The reported density also reproduces the proposed Coulomb phase:

\begin{equation}
\Omega_p^2=\frac{e^2n_e}{\epsilon_0m\gamma_0^3},\qquad\mu_Q=\frac{\Omega_pL}{c}\simeq0.145.
\label{eq:141}
\end{equation}

This is a collective-response parameter, not an rms energy width to add in quadrature. Its gain-dilution role
can resemble inhomogeneous broadening \cite{ref37}; the actual induced chirp requires the bunch field and
geometry. The broader high-gain comparison is given in Appendix A.

\subsection*{Divergence, overlap and intensity-dependent redistribution}

External focusing can place a waist inside an otherwise unfocused interaction region. The quoted emittance and
radius give, neglecting collective forces,

\begin{equation}
\sigma_{x'}=\frac{\epsilon_n}{\beta\gamma_0\sigma_e}\simeq0.233\,\mathrm{mrad},\quad\beta_*\simeq\frac{\gamma_0\sigma_e^2}{\epsilon_n}=2.92\,\mathrm{mm},\quad\sigma_x(z)=\sigma_e\sqrt{1+z^2/\beta_*^2}.
\label{eq:142}
\end{equation}

For a centred waist, the size grows by only $1.9\%$ at either end of the interaction. Free drift preserves the
angular distribution; it develops position--angle correlations rather than generating additional angular
spread. A density-weighted hourglass estimate gives $H_{e}=(2\beta_{\ast}/L)\arctan
[L/(2\beta_{\ast})]\simeq0.988$. This is not a universal gain correction: the actual electron--optical-mode
overlap, including the stored X-ray mode, must be evaluated.

Angular detuning is more significant. In the article's optical-undulator convention, for a fixed axial
radiation mode its equivalent electron-energy detuning is approximately

\begin{equation}
d_\theta=\frac{\gamma_0^2(x'^2+y'^2)}{2(1+a_0^2)},\qquad a_0=0.0944.
\label{eq:143}
\end{equation}

For independent equal Gaussian angles in the two planes, its mean and standard deviation are both about
$1.44\times10^{-4}$. Retuning can remove the mean, but not the angular distribution. Combining its variance
with independent entrance energy variance gives a diagnostic width $1.62\times10^{-4}\simeq2.2\Gamma$. The
angular distribution is asymmetric, so the full joint average, rather than a Gaussian width alone, is
required. These estimates identify a demanding acceptance margin; they do not prove that recoil is unresolved
or that the proposed state cannot be reached.

The growing intracavity field also changes the outgoing electron distribution. Ordinary low-gain oscillator
theory gives an induced relative energy spread proportional to the square root of local stored intensity
\cite{ref52}; its small-signal expression must not be extrapolated uncritically to the recoil-resolved regime.
In the KG model, stimulated recoil back-action is already included in Eq. (136). For an initially sharp
resonant electron, the two outgoing energy populations yield

\begin{equation}
\sigma_{\epsilon,\rm ind}=r\sqrt{P_{\rm em}(1-P_{\rm em})}.
\label{eq:144}
\end{equation}

Fresh bunches do not carry this outgoing spread from one round trip to the next, but each bunch experiences
the field accumulated in the cavity. The two-state approximation must remain valid throughout that buildup.

For their near-threshold example, $\delta=0.01$ gives

\begin{equation}
n_{\rm ss}\simeq\frac{3\delta N_e}{G}=2.07\times10^7,\qquad N_{\rm rt}\simeq\frac{\ln(3N_e\delta/G^2)}{G\delta}\simeq1.9\times10^4.
\label{eq:145}
\end{equation}

At 10 MHz this is approximately $1.9$ ms. The resonant estimate $gT\sqrt{n_{\rm
ss}}\simeq\sqrt{3\delta}=0.173$ gives $P_{\mathrm{em}}\simeq0.030$ and
$\sigma_{\epsilon,ind}\simeq8.0\times10^{-5}$, comparable with the entrance spread and acceptance, but below
the recoil. Near-threshold equilibrium therefore need not imply deep saturation or recoil washout. It remains
sensitive to beam averaging and optical overlap; the quoted buildup time and equilibrium should be reassessed
with those effects.

For independent contributions variances add, while correlated contributions also require covariance terms.
Adding the illustrative entrance, angular and outgoing induced widths in quadrature gives about
$1.8\times10^{-4}$, or $2.4\Gamma$. This is only a comparison of scales: the induced distribution is
conditional on the input and the field, and inserting that outgoing width back into the entrance integral
would double-count stimulated dynamics. A joint electron--field calculation should propagate the
redistribution once. The present scale estimates therefore motivate refinement rather than a categorical
negative conclusion.

\subsection*{Experimental signature and constructive perspective}

The decisive test is photon-number statistics in a defined detected mode or pulse. Calibrated photon counting,
or an equivalent calibrated pulse-energy measurement, can determine $F$. For independent loss and detection
with net efficiency $\eta$,

\begin{equation}
F_{\rm det}=1+\eta(F_{\rm cav}-1).
\label{eq:146}
\end{equation}

Loss moves $F$ toward unity; technical fluctuations, detector response and multimode collection must be
characterized separately. Sub-Poissonian counting would establish nonclassical radiation, whereas reduced
super-Poissonian fluctuations alone would not. Attribution to recoil-selective dynamics requires scans of
detuning, energy spread, divergence and stored intensity against compatible quantum and classical predictions.
A very weak interaction minimizes induced redistribution but may also weaken the statistical signal; the
useful experimental window must satisfy both gain--loss balance and observability.

The proposal's importance lies in opening a concrete route to such dedicated tests, rather than in already
proving experimental observability. Recoil separation is plausible for the reported entrance spread, but the
combined beam-quality and intensity-dependent effects leave a demanding resonance margin. These criticalities
should encourage further calculations and controlled experiments. The authors' innovative oscillator approach
is a positive contribution toward separating genuine quantum signatures from classical beam and cavity
fluctuations.

\subsection*{Kapitza--Dirac diffraction and a complementary coincidence experiment}

The Kapitza--Dirac effect offers a direct physical bridge between matter-wave diffraction and the microscopic
exchange underlying an optical FEL. Kapitza and Dirac proposed electron scattering by standing light waves
\cite{ref53}; resolved electron diffraction was later demonstrated by Freimund, Aflatooni and Batelaan
\cite{ref54}, with the development and interpretation reviewed by Batelaan \cite{ref55}. Its importance here
is that a periodic optical interaction produces discrete electron momentum orders. This establishes quantum
electron diffraction, while a sufficiently strong prescribed classical optical potential can already describe
that diffraction. It does not alone establish nonclassical radiation or quantum FEL amplification.

A complementary development of the Kling--Giese experimental perspective is a cavity supporting two quantized
counterpropagating optical modes, crossed transversely by individually detected electrons. Let the light
propagate along $\pm z$, the electron transport predominantly along $x$, and $p_{0}$ denote the entrance
$z-$momentum, scanned by a small incidence-angle change. In an effective nonrelativistic ponderomotive
reduction,

\begin{equation}
H_{\rm KD,int}=\hbar g\left(a_-^\dagger a_+e^{2ikz}+a_+^\dagger a_-e^{-2ikz}\right),\qquad q=2\hbar k.
\label{eq:147}
\end{equation}

The two exchange directions correlate $p_{0}\to p_{0}\pm q$ with a photon transferred into the opposite
travelling optical mode. The small-angle electron deflection is approximately $\Delta\theta=\pm q/p_{x}$. A
ring cavity with independently accessible travelling modes is a candidate geometry; output-mode mixing, cavity
backscattering and delayed leakage must be calibrated before photon counts can be assigned to directional
transfer. The relativistic implementation requires the appropriate reduction of the laboratory Hamiltonian
rather than direct use of the nonrelativistic coupling.

The proposed diagnostic is electron deflection in coincidence with the optical outcome. If $P_{+}(p_{0})$ and
$P_{-}(p_{0})$ are the weak single-exchange probabilities for increasing and decreasing the $-$ mode
occupation, respectively, the mean transfer is

\begin{equation}
\mathcal A_-(p_0)\equiv\langle n_-\rangle_{\rm out}-\langle n_-\rangle_{\rm in}=P_+(p_0)-P_-(p_0),\qquad\mathcal G_-(p_0)=\frac{\mathcal A_-(p_0)}{\langle n_-\rangle_{\rm in}}.
\label{eq:148}
\end{equation}

The first equality to the probability difference is a single-exchange approximation. $\mathcal{A}_{-}$ is the
mode-transfer gain curve; fractional gain $G_{-}$ requires a nonzero seed, whereas vacuum input gives a photon
yield. For the ideal interaction, conservation of $p_{z}-qn_{-}$ yields the exact balance $\Delta\langle
p_{z}\rangle=q\Delta\langle n_{-}\rangle$, before leakage or other momentum-changing processes. With equal
frequencies and symmetrically populated input modes, the two directional probabilities are interchanged by
$p_{0}\to-p_{0}$. The mean transfer is then antisymmetric and vanishes at $p_{0}=0$, even when both scattering
channels are populated. For signed exchange order $\ell=+1,-1,0$,

\begin{equation}
\langle\ell\rangle=P_+-P_-,\qquad\langle\ell^2\rangle=P_++P_-.
\label{eq:149}
\end{equation}

Coincidences and fluctuations can thus reveal microscopic exchange hidden by a zero mean. The averaging issue
identified by Carmesin et al. \cite{ref43} supplies the motivation for this diagnostic: agreement of first
moments is a weaker test than agreement of joint outcomes. The measured transfer still requires convolution
with the entrance momentum distribution; conditional detection does not remove unresolved beam uncertainty.

For equal optical frequencies the ideal exchange conserves total optical energy: one mode grows at the expense
of the other. Electron-powered optical amplification requires a suitable frequency difference and Doppler
resonance. Low mode occupation limits perturbation, but must coexist with sufficient single-photon coupling
and detection efficiency. An incoming number state is already a nonclassical resource; a weak coherent-state
control is needed to distinguish transferred input nonclassicality from effects generated in the interaction.
Directional population correlations identify exchange but do not by themselves certify entanglement, which
requires a coherence-sensitive witness. The proposal is therefore a constructive route to controlled
electron--photon quantum dynamics, informed by Carmesin's averaging analysis and Walther's
preparation-and-readout strategy, with experimental feasibility still to be established.

\subsection*{Stimulated \v{C}erenkov exchange as an experimental precedent}

The observation of stimulated quantum Cherenkov exchange by Nehemia and collaborators provides an encouraging
experimental precedent \cite{ref56}. Electrons grazing a dielectric prism interact with the evanescent optical
field produced by total internal reflection. Matching the longitudinal optical phase velocity to the electron
velocity permits coherent interaction over hundreds of micrometres, producing resolved electron energy
sidebands separated by $\hbar\omega$. At a representative kinetic energy of $207.2$ keV and optical wavelength
of 730 nm, the photon spacing is approximately $1.7$ eV. The reported spectrometer resolution is approximately
$0.1$ eV; importantly, the effective resolution of individual sidebands is limited by the electron zero-loss
peak width, approximately $1.1$ eV in most measurements. These results demonstrate that adequate preparation,
synchronism and energy resolution can make stimulated quantum electron dynamics experimentally accessible.

This perspective complements the Kapitza--Dirac and cavity proposals discussed above. The \v{C}erenkov
geometry supplies longitudinal synchronism through the dielectric optical field, whereas the proposed two-mode
Kapitza--Dirac experiment measures transverse momentum exchange. Resolved electron sidebands support the
quantum description of the electron, but can also be obtained with a prescribed coherent optical field; their
observation does not by itself establish nonclassical photon statistics or recoil-resolved FEL oscillation.
Extending such measurements to electron--photon coincidences would therefore address a further question. The
reported sensitivity of the interaction strength to electron energy, and the trade-off between interaction
length and angular spreading near the surface, reinforce the importance of characterized beam preparation. The
achievement nevertheless supports the constructive conclusion that dedicated geometries and suitable
sensitivity can expose microscopic quantum exchange.

The connection with \v{C}erenkov FELs also has an established theoretical background. Walsh, Johnson, Dattoli
and Renieri compared undulator and \v{C}erenkov free-electron lasers in the Compton regime for short-pulse
relativistic electron beams \cite{ref57}. This comparison places the present experimental perspective within
the broader study of stimulated electron--radiation exchange, while retaining the distinct synchronism
mechanisms of the two devices.

\section{Conclusions}

The analysis shows that recoil observability depends on accumulated transition phases, electron and optical
preparation, and the acceptance of the measured interaction. Neither spontaneous startup nor excess photon
variance alone establishes a nonclassical FEL signature.

\subsection*{Cumulative recoil and the Bloch--Nordsieck interpretation}

The connection with Bloch and Nordsieck \cite{ref32} should be understood at the level of the
prescribed-current, no-recoil approximation. An individual photon transfer may be small while its accumulation
over many transfers changes the electron motion appreciably. Accordingly, the controlling quantity involves
both the elementary recoil and the populated exchange orders. This is a physical restriction on the
approximation, rather than a consequence of the infrared cancellation theorem itself.

In the notation of this article, $\varepsilon$ is the dimensionless Raman--Nath recoil coefficient of Eq.
(46), not the fractional laboratory photon energy. The neighboring-state mismatch of Eq. (50) is

\begin{equation}
\delta_\ell^\pm=\varepsilon(2\ell\pm1),\qquad|\delta_\ell^\pm\tau|\ll1.
\label{eq:150}
\end{equation}

The small-phase condition must hold over the significantly occupied ladder throughout the interaction. For
large occupied $|\ell|$, its characteristic dependence is $|\varepsilon\ell\tau|\ll1$, up to a factor of two.
At a fixed normalized exit time of order unity this is the origin of the useful shorthand
$|\varepsilon\ell|\ll1$. Comparing a recoil shift with a resonance acceptance also requires both quantities in
the same frequency or detuning normalization; $\varepsilon\ell$ cannot be compared directly with the Pierce
parameter without that conversion.

For a static magnetic undulator, the fractional laboratory energy transfer per resonant radiation photon is

\begin{equation}
\eta_{\rm rec}=\frac{\hbar\omega_L}{E_e},\qquad E_e=\gamma mc^2,\qquad\frac{\Delta E_e}{E_e}\simeq-\ell\eta_{\rm rec}.
\label{eq:151}
\end{equation}

Here $\ell$ counts net photons emitted by the electron into the selected mode, assuming a nearly fixed photon
frequency and a small relative energy change. For purely emissive transfers it is the emitted photon count;
with emission and absorption it is the signed exchange count. It is not the total photon occupation of the
cavity. In an optical-pump geometry the electron energy transfer must instead include the pump-photon exchange
consistently with the Hamiltonian and chosen frame. Neglecting the accumulated mean energy loss relative to
the electron energy requires

\begin{equation}
|\ell|\eta_{\rm rec}\ll1.
\label{eq:152}
\end{equation}

This condition is stronger than requiring $\eta_{\mathrm{rec}}\ll1$ alone. For a distribution of exchange
orders, one must also control the appreciably populated tails and the induced width: cancellation of the mean
signed exchange does not imply negligible recoil fluctuations or negligible electron--field correlations.

The Bloch--Nordsieck analogy is that an arbitrarily large photon population can accompany a small total
radiative energy and momentum transfer. In the infrared spectrum, $dN\propto d\omega/\omega$, whereas
$dE=\hbar\omega dN\propto d\omega$. Photon number may therefore diverge without an energy divergence
\cite{ref32}. Neglecting the finite accumulated transfer is an additional dynamical approximation. Radiation
from a prescribed accelerated trajectory remains possible even when its radiation reaction is negligible.

\subsection*{Bandwidth, saturation and the scope of the approximation}

For FEL resonance, the relevant comparison is more restrictive than the total relative energy criterion above.
In conventional high-gain scaling, the electron energy acceptance is of order $\rho_{1}E_{e}$. A negligible
change of resonance due to the accumulated mean exchange requires, as a scale estimate,

\begin{equation}
\frac{|\ell|\eta_{\rm rec}}{\rho_1}\ll1.
\label{eq:153}
\end{equation}

Order-unity coefficients depend on the adopted bandwidth and detuning conventions. This mean-shift criterion
supplements the local phase condition and the distribution-width comparisons; it does not replace them. For
example, a small mean exchange can coexist with an induced energy spread comparable with the resonance
acceptance.

The characteristic number of selected-mode photons transferred per electron when a fraction $\chi$ of its
energy is extracted is

\begin{equation}
\ell_{\rm sat}\sim\frac{\chi E_e}{\hbar\omega_L}=\frac{\chi}{\eta_{\rm rec}},\qquad\ell_{\rm sat}\eta_{\rm rec}\sim\chi=O(\rho_1).
\label{eq:154}
\end{equation}

This is an energy-balance estimate for the average net transfer, not a statement that each electron emits the
same number of photons. For radiation power $P_{\mathrm{rad}}$ and electron beam power $P_{E},\chi=$
$P_{\mathrm{rad}}/P_{E}$. The booklet convention used above gives $\chi_{\mathrm{sat}}=2\rho_{1}$ at
$P_{F,1}=2\rho_{1}P_{E}$; the frequently used estimate $\chi_{\mathrm{sat}}\sim\rho_{1}$ has the same
parametric content.

Thus the accumulated energy loss at saturation is small relative to the beam energy when $\rho_{1}\ll1$, yet
is of the order of the FEL resonance acceptance. A fixed-trajectory approximation that suppresses this
back-action cannot describe the approach to saturation self-consistently. Classical FEL theory retains the
continuous evolution of electron energy, phase, bunching and field amplitude, while it can neglect the
discreteness of the elementary quantum transfers.

The distinction is particularly clear using the parameter already defined in the conclusions,

\begin{equation}
\bar\rho=\frac{\rho_1}{\eta_{\rm rec}},\qquad\ell_{\rm sat}=O(\bar\rho),\qquad\frac{\ell_{\rm sat}\eta_{\rm rec}}{\rho_1}=O(1).
\label{eq:155}
\end{equation}

When $\bar{\rho}\gg1$, many elementary transfers fit within the characteristic extraction scale and a
continuous classical description can be appropriate. Their cumulative effect nevertheless remains essential.
When $\bar{\rho}$ approaches unity or below, recoil discreteness can become dynamically important, subject to
the beam preparation, coherence and diffusion limitations discussed in this article.

Finally, resonant FEL photons occupy a finite frequency band and their emission involves stimulated exchange
and collective feedback. Negligible accumulated recoil therefore does not establish an infrared soft-photon
limit. The precise conclusion is that a Bloch--Nordsieck-type no-recoil treatment is justified only when the
total radiation-induced change in electron motion is negligible on the scales and over the duration relevant
to the calculation. The inclusive infrared cancellation theorem is broader and does not require that total
energy transfer vanish. This interpretation reconciles the usefulness of the prescribed-current reference with
the necessity of retaining electron back-action in a working FEL.

Section 7 of Ref. \cite{ref27} establishes the historical distinction between joint exchange-ladder coherence,
recoil-induced statistical corrections and experimentally measured radiation coherence. Its warning about
realistic beam distributions and longitudinal modes remains central to the observability problem. The
prescribed-current source is gainless in its fixed-trajectory stimulated balance; a self-consistent classical
electron trajectory can nevertheless yield gain with coherent conditional radiation.

The minimal-coupling and interaction-picture derivations identify detuning, recoil and photon exchange within
one consistent dynamical framework. The prescribed-current solution supplies a coherent Poisson reference,
whereas a quantum electron can become correlated with the field. The Raman--Nath equation makes recoil phases
and their perturbative correction explicit, provided the populated ladder and its validity range are
specified.

The physical control is the neighboring-state mismatch. Small $\varepsilon$ alone is insufficient if
significantly occupied recoil orders accumulate large phases. A second limitation arises from the electron
distribution: field-induced modulation and energy spread generate phase mixing in low gain as well as in
collective amplification. Beam convolution supplies an initial response but does not replace self-consistent
electron--field evolution.

The Wigner treatment defines the Moyal bracket and derives the finite momentum translation from the sinusoidal
interaction. It distinguishes recoil populations from coherences responsible for spatial bunching. Smoothness
on the recoil scale, accumulated recoil phase and beam dephasing are complementary criteria, rather than
interchangeable measures of quantumness.

The oscillator comparison then embeds single-pass dynamics in a fresh-electron channel and cavity loss. Kling
and Giese provide a recoil-resolved realization with many effective optical-undulator periods despite its
short laboratory length and, in suitable nonlinear operating regions, sub-Poissonian photon statistics. This
complements the perturbative Raman--Nath regime. The appendices distinguish low-gain quantum passages from
high-gain SASE and explain how discrete-electron noise seeds collective amplification. They supply a dynamical
comparison and a statistical baseline rather than independent evidence of observable quantum recoil.

Three competing effects must be compared quantitatively before interpreting a small quantum signature:
stimulated exchange with the amplified selected mode, spontaneous emission into other radiation modes, and the
self-consistent FEL-induced electron distribution change. The first and third are descriptions of the same
selected-mode back-action in different regimes; their variances must not be added as independent sources.
Spontaneous emission into unobserved modes is a separate contribution. Small initial energy spread, low gain
per passage and small signal are not, individually, sufficient guarantees of observability. Bunch charge,
duration and transverse size must also be specified: their collective Coulomb response can change gain and
generate phase-dependent energy modulation. In the approximate high-gain equation of Ref. \cite{ref37}, space
charge and recoil enter through additive squared parameters, so intensity evolution alone does not disentangle
them. The reported entrance spread and acceptance are compared in Eq. (139); Appendix A separately analyzes
correlated Coulomb modulation and the approximate high-gain space-charge/recoil degeneracy.

For the KG oscillator the stimulated contribution must be calculated with the stored cavity radiation,
including photons retained from earlier passages. For a sharp upper-state electron momentum at resonance, the
weak-transition probability is

\begin{equation}
P_{\rm em}(n)\simeq(gT)^2(n+1)=(gT)^2+(gT)^2n.
\label{eq:156}
\end{equation}

The first term is spontaneous emission into the selected empty mode and the second is stimulated emission. To
leading order the corresponding stimulated momentum variance is

\begin{equation}
\sigma_{p,\rm stim}^2\simeq q^2(gT)^2\overline n_{\rm cav},\qquad\sigma_{p,\rm ind}^2\simeq q^2(gT)^2(\overline n_{\rm cav}+1).
\label{eq:157}
\end{equation}

The mean $\bar n_{\rm cav}$ must be obtained from pumping and cavity loss, not replaced by a vacuum input on
every passage. More generally, in the resonant two-state model,

\begin{equation}
\overline P_{\rm em}=\sum_n P_n^{\rm cav}\sin^2\!\left(gT\sqrt{n+1}\right),\qquad\sigma_{p,\rm ind}^2=q^2\overline P_{\rm em}(1-\overline P_{\rm em}).
\label{eq:158}
\end{equation}

The variance formula assumes an initially sharp electron momentum; finite-width inputs require the joint
momentum--transition average and covariance of Eq. (89). Photon-counting probabilities suffice for this ideal
upper-state input, whereas a general electron preparation may also require coherence information.

Let $\xi=gT\sqrt{\bar n_{\rm cav}}$. Comparing the stimulated rms spread with the finite-time momentum
acceptance discussed in Section XII gives the scale estimate

\begin{equation}
\frac{\sigma_{p,\rm stim}}{q/(\omega_rT)}\sim\xi\omega_rT.
\label{eq:159}
\end{equation}

Consequently, $\xi\ll1$ need not make the induced spread negligible compared with a narrow recoil-selective
acceptance when $\omega_{r}T\gg1$. This comparison does not itself prove that resonance is destroyed: the
ideal two-branch variance includes the intended recoil transition, not merely incoherent broadening within a
branch. Fresh electrons prevent accumulation of the outgoing spread in the next bunch, but the stored field
still controls each new passage. Branch widths, additional recoil orders and coherence must be checked at the
actual intracavity intensity. Increasing that intensity also increases $g\sqrt{n+1}/\omega_{r}$, eventually
invalidating the two-state approximation. Low gain therefore does not mean absence of stimulated electron
back-action.

For high-gain SASE, even an initially cold beam accumulates uncorrelated spread from spontaneous undulator
radiation. Define $D_{\gamma}=d\sigma_{\gamma,sp}^{2}/dz$, with $\sigma_{\gamma}$ the rms spread of the
Lorentz factor. In an ideal planar magnetic undulator, with peak deflection parameter $K$, the standard
diffusion expression is \cite{ref58}

\begin{equation}
D_\gamma=\frac7{15}r_e\bar\lambda_C\gamma^4 k_u^3 K^2 F(K),\qquad F(K)=1.20K+\frac1{1+1.33K+0.40K^2}.
\label{eq:160}
\end{equation}

Here $r_{e}$ is the classical electron radius, $\bar{\lambda}_{C}=\hbar/(mc)$, and
$k_{u}=2\pi/\lambda_{u};D_{\gamma}$ has units $\mathrm{m}^{-1}$. This uncorrelated diffusion has been measured
at the European XFEL. The formula is for a magnetic undulator, not a direct substitution rule for the KG
optical pump. Let $\Delta\gamma_{\mathrm{rec}}=\hbar\omega_{L}/(mc^{2})$ be the laboratory energy change
associated with one selected-mode radiation photon in the static-undulator case. The accumulated comparison is

\begin{equation}
\sigma_{\gamma,\rm sp}^2(L)=\int_0^L D_\gamma(z)\,dz,\qquad\mathcal R_{\rm diff}(L)=\frac{\sigma_{\gamma,\rm sp}^2(L)}{(\Delta\gamma_{\rm rec})^2}.
\label{eq:161}
\end{equation}

For constant parameters and fundamental planar resonance,

\begin{equation}
\mathcal R_{\rm diff}=\frac{7\pi}{30}\alpha_{\rm fs}N_uK^2F(K)\left(1+\frac{K^2}{2}\right)^2,\qquad L=N_u\lambda_u,
\label{eq:162}
\end{equation}

where $\alpha_{\mathrm{fs}}=r_{e}/\bar{\lambda}_{C}$. For $K=1$ and $N_{u}=1000$ this gives
$\mathcal{R}_{\mathrm{diff}}\simeq18.9$, or an rms diffusion spread about $4.3$ selected-mode recoils. This
calculated illustration shows why a small relative beam spread can exceed the microscopic recoil scale.

When the recoil term in a normalized high-gain dispersion relation is of order
$[\Delta\gamma_{\mathrm{rec}}/(\rho\gamma)]^{2}$, while a narrow-distribution spread correction is of order
$[\sigma_{\gamma}/(\rho\gamma)]^{2}$, their scale ratio is Eq. (161). Exact coefficients and even the
applicability of a variance expansion depend on the particular gain equation and distribution. Diffusion small
compared with the ordinary FEL bandwidth can nevertheless exceed the small quantum correction being sought.
The comparison must cover the amplification length required to observe the effect. It is not sufficient to
characterize the entrance beam alone. In a drift--diffusion approximation the spontaneous contribution enters
as $(D_{\gamma}/2)\partial_{\gamma}^{2}f$ in the longitudinal kinetic evolution, together with mean radiative
energy loss. For rare large photon jumps an emission jump operator is required instead; the full spontaneous
spectrum determines its moments \cite{ref59}.

The stimulated high-gain broadening is quantified by the design booklet's Eqs. (3.11)--(3.12) \cite{ref60}.
Denoting relative rms spreads by $\sigma_{i}$ and $\sigma_{\varepsilon}$, those formulas read

\begin{equation}
\begin{aligned}
\sigma_i(z)&\simeq3C\sqrt{\frac{A(z)}{1+9B[A(z)-1]}},& C&=\frac12\sqrt{\frac{\rho_1P_0}{P_E}},\\
B&\simeq\frac{1.24P_0}{9P_{F,1}},& \sigma_\varepsilon^2(z)&=\sigma_{\varepsilon,0}^2+\sigma_i^2(z),\quad\sigma_{i,F}\simeq1.6\rho_1.
\end{aligned}
\label{eq:163}
\end{equation}

Here $P_{E}$ is the beam power, $P_{0}$ the input radiation power, $\rho_{1}$ the fundamental Pierce parameter
and $P_{F,1}=2\rho_{1}P_{E}$ the booklet's saturation power. Its linear growth function is

\begin{equation}
A(z)=\frac19\left[3+2\cosh\left(\frac z{L_{g,1}}\right)+4\cos\left(\frac{\sqrt3 z}{2L_{g,1}}\right)\cosh\left(\frac z{2L_{g,1}}\right)\right].
\label{eq:164}
\end{equation}

These are approximate classical design formulas, not an exact initial-value solution for arbitrary quantum
preparations. In the exponential interval before saturation, using $P(z)\simeq P_{0}A(z)$ and neglecting the
slowly varying fitting factor in the booklet's power formula, they imply

\begin{equation}
\sigma_i^2(z)\simeq\frac94\rho_1\frac{P(z)}{P_E}.
\label{eq:165}
\end{equation}

An effective startup power may be used for a SASE scaling estimate, but a fluctuating multimode SASE field
requires its own statistical evolution rather than identification with a deterministic seed.

Define $\eta_{\mathrm{rec}}=\hbar\omega_{L}/E_{e}$ and $\bar{\rho}=\rho_{1}/\eta_{\mathrm{rec}}$, where
$E_{e}=\gamma mc^{2}$. The stimulated variance relative to the recoil-squared scale is then

\begin{equation}
\mathcal R_{\rm FEL}(z)=\frac{\sigma_i^2(z)}{\eta_{\rm rec}^2}\simeq\frac94\bar\rho^2\frac{P(z)}{\rho_1P_E}.
\label{eq:166}
\end{equation}

For a small additive recoil correction of order $1/\bar{\rho}^{2}$, stimulated spread can exceed this scale
when

\begin{equation}
\frac{P(z)}{P_{F,1}}\gg\frac{2}{9\bar\rho^2}.
\label{eq:167}
\end{equation}

For example, at $\bar{\rho}=100$, equality occurs at $P/P_{F,1}\simeq2.2\times10^{-5}$, well below saturation.
The practical obstacle is therefore not only externally supplied energy spread: the amplification needed to
reveal a small correction can itself generate a larger change in the electron distribution.

This comparison must retain its dynamical meaning. FEL-induced spread is correlated with electron phase,
bunching and field growth. It cannot generally be substituted for a fixed Gaussian entrance spread or an
independent diffusion coefficient. In particular, inserting Eq. (163) into a cold-beam cubic gain equation is
a scale estimate, not a self-consistent derivation. The evolving inhomogeneous response can require a kinetic
or integral formulation. Likewise, a variance exceeding the recoil-squared scale does not alone prove that
every quantum observable is erased: one must compare the actual observable calculated with stimulated
evolution, spontaneous diffusion and the compatible beam preparation. The classical booklet estimates support
a serious limitation for small perturbative quantum corrections; they do not establish a universal exclusion
of the recoil-resolved regime $\bar{\rho}\lesssim1$, where a quantum self-consistent treatment replaces those
classical estimates.

A meaningful experimental claim must show that its specified recoil or correlation observable survives
averaging, technical fluctuations, loss and mode selection under achievable beam and undulator conditions.
Neither startup alone nor narrowing relative to another broad distribution establishes that claim.

This requirement leaves a constructive experimental direction. Carmesin and collaborators identify how
microscopic quantum structure and averaged gain can differ; Kapitza--Dirac exchange offers a controllable
momentum-space interaction; and Walther's micromaser research supplies a precedent for particle preparation
and cavity readout. A dedicated few-electron, low-occupation experiment with deflection--photon coincidences
can bring these perspectives together. The aim is to test specified joint observables under characterized
conditions, while establishing the coupling and coherence needed for an eventual quantum FEL experiment.

Recoil observability therefore depends on accumulated transition phases, electron and optical preparation, and
the acceptance of the measured interaction. Neither spontaneous startup nor excess photon variance alone
establishes a nonclassical FEL signature.

\section*{Acknowledgments and AI-use disclosure}

During the preparation of this article, the authors used OpenAI's ChatGPT, including the Codex assistant in
Work mode, to assist with calculations, checks of derivations, literature searches, numerical cross-checks,
and revision of the English text. The model versions used at different stages of the work were not
systematically recorded; the final revision was checked using GPT-5.6 Sol.

The authors conceived and directed the study, drawing on their earlier work on the quantum theory of
free-electron lasers and on subsequent contributions by other researchers. AI-assisted calculations were used
to check analytical results and to compare calculations performed with Mathcad and Mathematica. AI tools also
assisted in locating relevant publications and in examining concise derivations in the literature. The cited
publications were consulted by the authors.

All scientific interpretations, approximations, analytical and numerical results, and the final wording were
reviewed by the authors, who take full responsibility for the content of the article.

\appendix
\renewcommand{\theequation}{\Alph{section}.\arabic{equation}}

\setcounter{equation}{0}

\section{Low-gain quantum FEL dynamics and high-gain SASE}

\subsection*{Different dynamical problems}

The low-gain quantum description resolves joint momentum--photon exchange during one passage. An incoming
electron state and optical state specify a unitary transition kernel; electron tracing and entrance averaging
then give observable probabilities. Low gain does not imply a single-electron beam: an independent-passage
reduction can average over many electrons, while a bunch with appreciable common-field correlations requires
its full channel. In an oscillator, fresh-electron pumping and losses evolve the stored field over successive
passages.

High-gain SASE is a many-electron traveling-wave initial-value problem. A finite beam supplies a stochastic
source; the common field changes electron phases and energies, which change the source in turn. Slippage,
longitudinal modes and the evolving distribution matter. A sum of independent emission probabilities cannot
replace this feedback. Recoil can be included in a self-consistent many-electron ladder or density-matrix
description \cite{ref9,ref16}; its classical limit retains collective dynamics while smoothing elementary
transfers.

\begin{center}
\begin{tabular}{@{}p{.16\linewidth}p{.37\linewidth}p{.37\linewidth}@{}}
Aspect & Low-gain passage / oscillator & High-gain SASE\\ \hline
Field evolution & Small change per pass; stored field evolves with pumping and loss & Traveling field and electron distribution evolve together\\
Electron input & Prepared state or mixture; fresh electrons in the cavity map & Beam distribution, finite-electron noise and longitudinal structure\\
Quantum test & Recoil transitions, conditional correlations, cavity photon statistics & Recoil-dependent gain or correlations surviving collective growth\\
Limitation & Transit acceptance, branch coherence and stored-field mixing & Entrance quality, diffusion, induced modulation, slippage and modes
\end{tabular}
\end{center}

A conditional low-gain Poisson law does not establish SASE statistics. Classical high-gain bunching and growth
do not alone identify recoil. Both regimes require propagation of compatible entrance information to the same
measured observable.

\subsection*{Space charge, energy modulation and the high-gain response}

The following discussion concerns a collective beam and its high-gain response. It supplies a broader
observability comparison; it is not a quantitative correction to the low-gain Kling--Giese two-state cavity
model. Charge-density and plasma-frequency quantities were defined in Eqs. (138)--(141). The distinction
between correlated Coulomb chirp, uncorrelated entrance spread and spontaneous diffusion is essential when
using an intensity-based gain equation.

Space charge redistributes energy and phase, and can generate a correlated longitudinal chirp. It is not a
photon-emission diffusion coefficient. A simple estimate makes the dependence on geometry explicit: for a cold
uniform ellipsoid of laboratory density $n_{e}$, let $D_{z}$ be the longitudinal depolarization factor of its
rest-frame shape. The linear internal electric force in laboratory coordinates is $F_{z}\simeq
e^{2}n_{e}D_{z}z/\epsilon_{0}$. Over a short interval with negligible expansion,

\begin{equation}
\sigma_{\gamma,\rm SC}\sim\frac{\beta e^2n_eD_zT_{\rm lab}\sigma_z}{\epsilon_0mc},\qquad\sigma_{p,\rm SC}\sim\frac{e^2n_eD_zT_{\rm lab}\sigma_z}{\epsilon_0}.
\label{eq:A.1}
\end{equation}

These expressions are the rms width of the induced linear chirp, not an independent variance to add blindly to
an already correlated distribution. The estimate follows from $\Delta p_{z}=F_{z}T_{\mathrm{lab}}$ and
$\Delta\gamma\simeq$ $\beta\Delta p_{z}/(mc)$. It requires $\Omega _{p}T_{\mathrm{lab}}\ll1$ for the
frozen-shape approximation and excludes acceleration, conducting boundaries and external compensation. The
uniform-ellipsoid density is not the Gaussian central density of Eq. (138). The factor $0<D_{z}<1$ depends on
the transformed aspect ratio; a Gaussian bunch or a beam in conducting boundaries requires its actual Poisson
field. Initial correlations, focusing and expansion must be propagated for a quantitative result. The induced
width must be compared with the recoil-equivalent energy width and resonance acceptance, including
slice-dependent detuning; a removable mean chirp and uncorrelated slice spread affect observations
differently.

The connection with the collective analysis of Ref. \cite{ref37} is particularly direct. In the notation and
approximate one-dimensional normalization of Eq. (37) of Ref. \cite{ref37},

\begin{equation}
a^{(3)}+2i\widetilde\phi a''+\left(\widetilde\mu_Q^2+\widetilde\mu_q^2-\widetilde\phi^2\right)a'=\frac{i}{3\sqrt3}a,\qquad\widetilde\mu_q=\frac{\hbar\omega_L}{\rho\gamma mc^2}=\frac1{\bar\rho}.
\label{eq:A.2}
\end{equation}

Primes are derivatives with respect to $\widetilde z=z/L_{g},L_{g}=\lambda_{u}/(4\pi\sqrt3\rho)$, and
$\widetilde\phi=(\omega_{0}-\omega)/(2\sqrt3\rho\omega_{0})$. The coefficient $\widetilde\mu_{Q}$ is the
space-charge parameter of that paper's high-gain model; its numerical normalization must be retained when
converting a physical plasma phase into the reported coefficient. This is the reference's phenomenological
recoil extension of a cold-beam equation, not a derivation of the full many-electron quantum hierarchy or a
formula to transplant unchanged into the optical cavity model.

Within that stated equation, the two effects enter only through
$M^{2}=\widetilde\mu_{Q}^{2}+\widetilde\mu_{q}^{2}$. An intensity evolution alone therefore cannot identify
their separate contributions. A comparison must independently determine the charge-density response and vary
charge, current or geometry under controlled beam conditions. Increasing recoil or reducing entrance energy
spread does not remove this degeneracy. On the same coefficient scale the ratio is
$\widetilde\mu_{Q}^{2}/\widetilde\mu_{q}^{2}$; collective Coulomb response can exceed the proposed recoil
correction even while both perturb ordinary gain only weakly.

For the homogeneous model with $a(0)$ prescribed and $a^{\prime}(0)=a^{\prime\prime}(0)=0$, the equivalent
memory equation is

\begin{equation}
\frac{da}{d\widetilde z}=\frac{i}{3\sqrt3}\int_0^{\widetilde z}e^{-i\widetilde\phi s}\frac{\sin(Ms)}M\,a(\widetilde z-s)\,ds,\qquad\frac{\sin(Ms)}M\longrightarrow s\quad(M\to0).
\label{eq:A.3}
\end{equation}

It follows by factoring the operator acting on $a^{\prime}$ as $(\partial_{\widetilde
z}+i\widetilde\phi)^{2}+M^{2}$. A distribution of detunings modifies the kernel; a Gaussian static
distribution multiplies its phase factor by $e^{-\sigma_{\widetilde\phi}^2s^2/2}$. Inhomogeneous or evolving
beams generally retain a Volterra response rather than a universal constant-coefficient cubic. Ref.
\cite{ref37} stresses both the similarity of gain dilution and the heuristic limitations of its quantum
extension. A measured photon-statistical or electron--photon correlation observable must therefore be
calculated beyond an intensity-only comparison if it is to distinguish quantum information from collective
classical forces.

\setcounter{equation}{0}

\section{Shot noise and the early stages of FEL growth}

An exactly smooth deterministic Maxwell--Lorentz or Vlasov beam with a uniform phase distribution and zero
optical field can remain a symmetry-preserving solution. A real beam contains a finite number of electrons.
Their phases do not form a perfectly smooth distribution in a single realization. This discreteness gives
electron shot noise.

For $N$ independent phases $\theta_{j}$ uniformly distributed over $[0,2\pi)$,

\begin{equation}
b_N=\frac1N\sum_{j=1}^N e^{-i\theta_j},\qquad\langle b_N\rangle=0,\qquad\langle|b_N|^2\rangle=\frac1N.
\label{eq:B.1}
\end{equation}

The variance follows by separating the diagonal $j=k$ terms from off-diagonal terms whose independent phase
averages vanish. Thus the rms microscopic bunching is $N^{-1/2}$ with a random phase, rather than a nonzero
coherent ensemble mean.

The field couples to bunching and changes electron energies and phases. A random seed can excite an unstable
eigenmode of the coupled system. During a linear high-gain interval a component has the schematic form

\begin{equation}
a(\tau)\sim a_{\rm noise}e^{\lambda\tau},\qquad b(\tau)\sim b_{\rm noise}e^{\lambda\tau},\qquad\operatorname{Re}\lambda>0.
\label{eq:B.2}
\end{equation}

Shot noise supplies the seed; the instability of the coupled electron--field system supplies gain. For a
nonuniform beam or a time-dependent kernel the growth law need not reduce to a single constant exponential.
The expression is a local linear-mode illustration, not a universal cubic reduction.

If $A_{\mathrm{sat}}$ marks the end of the linear regime, then

\begin{equation}
\tau_{\rm sat}\sim\frac{1}{\operatorname{Re}\lambda}\log\left(\frac{A_{\rm sat}}{A_{\rm noise}}\right).
\label{eq:B.3}
\end{equation}

For a fixed gain rate, seed strength shifts startup only logarithmically. Changing the number of electrons can
also change current and gain, so the $N^{-1/2}$ seed scaling must not be used while silently holding all beam
dynamics invariant.

Three statements must be distinguished. A perfectly smooth deterministic zero-field beam does not create a
preferred phase by itself. A discrete classical beam radiates and can seed collective amplification without
quantizing the electromagnetic field. A fully quantum electron--field description contains vacuum fluctuations
and the $n+1$ emission factor, so emission is possible into an empty mode. Observing startup alone cannot
distinguish these descriptions.

Shot noise and recoil act at different levels. Electron discreteness changes the stochastic source; recoil
changes elementary transitions and their phase dependence. A realistic calculation must include both without
duplicating the same spontaneous feeding in two representations. The comparison should use an ensemble of
classical noise realizations propagated through the same collective dynamics and detection model, not merely
compare a quantum probability law with a smooth zero-field calculation.

Classical source and technical fluctuations can give super-Poissonian photon statistics. Within a
semiclassical mixture of coherent fields, $F=1+\Var (\mu)/\langle\mu\rangle\ge1$, where $\mu$ is the random
conditional mean. For a specified counting gate and collection of optical modes, statistically significant
sub-Poissonian photon counts are a stronger nonclassical diagnostic than excess variance or startup. The
measured counts must be related to the source through a calibrated detector response: independent losses move
$F$ toward unity, while background counts and electronic noise contribute their own means and variances.
Detector saturation or dead time can artificially suppress fluctuations and must be excluded. If several modes
are collected, the variance includes their individual variances and cross-covariances; the measured quantity
must be compared with a prediction for that same collection. These conditions specify how a reported $F<1$ is
assessed, rather than assuming an unspecified correction for ``background and multimode effects.''
Electron--photon correlations and conditional distributions can provide additional discriminants. The
shot-noise discussion supplies the classical statistical baseline; the phase-space study of Ref. \cite{ref43}
further illustrates why quantum structure need not survive observable averaging.


\begin{thebibliography}{60}

\bibitem{ref1} J. M. J. Madey, ``Stimulated emission of bremsstrahlung in a periodic magnetic field,'' Journal of Applied Physics 42, 1906--1913 (1971).

\bibitem{ref2} W. Becker and H. Mitter, ``Quantum theory of a free electron laser,'' Zeitschrift für Physik B 35, 399--404 (1979). DOI: 10.1007/BF01332701.

\bibitem{ref3} J. K. McIver and M. V. Fedorov, ``Quantum theory of stimulated processes in a free-electron laser in a strong field,'' Soviet Physics JETP 49, 1012--1019 (1979); Zh. Eksp. Teor. Fiz. 76, 1996 (1979).

\bibitem{ref4} G. Dattoli and A. Renieri, ``FEL Quantum Aspects,'' Journal de Physique Colloques 44(C1), C1-125--C1-136 (1983).

\bibitem{ref5} G. Dattoli and A. Renieri, The Quantum Mechanical Analysis of the Free Electron Laser, ENEA Frascati Report RT/TIB/85/42 (1985).

\bibitem{ref6} G. Dattoli, A. Renieri, F. Romanelli and R. Bonifacio, ``On the coherent states of a free electron laser,'' Optics Communications 34, 240--244 (1980).

\bibitem{ref7} G. Dattoli and A. Renieri, ``The quantum and classical aspects of the free electron laser theory,'' Optics Communications 39, 328--330 (1981).

\bibitem{ref8} A. Bambini and S. Stenholm, "Single-particle theory of the free-electron laser in a moving frame," IEEE Journal of Quantum Electronics QE-17, 1355 (1981).

\bibitem{ref9} W. Becker and J. K. McIver, ``Fully quantized many-particle theory of a free-electron laser,'' Physical Review A 27, 1030--1043 (1983). DOI: 10.1103/PhysRevA.27.1030.

\bibitem{ref10} W. Becker and J. K. McIver, ``Many-particle quantum theory for a class of free-electron devices,'' Physics Reports 154, 205--245 (1987). DOI: 10.1016/0370-1573(87)90068-8.

\bibitem{ref11} J. Gea-Banacloche, ``Quantum theory of the free-electron laser: Large gain, saturation, and photon statistics,'' Physical Review A 31, 1607 (1985). DOI: 10.1103/PhysRevA.31.1607.

\bibitem{ref12} A. Friedman, A. Gover, G. Kurizki, S. Ruschin and A. Yariv, ``Spontaneous and stimulated emission from quasifree electrons,'' Reviews of Modern Physics 60, 471--535 (1988). DOI: 10.1103/RevModPhys.60.471.

\bibitem{ref13} W. Becker and M. S. Zubairy, ``Photon statistics of a free-electron laser,'' Physical Review A 25, 2200--2207 (1982). DOI: 10.1103/PhysRevA.25.2200.

\bibitem{ref14} W. Becker, M. O. Scully and M. S. Zubairy, ``Generation of squeezed coherent states via a free-electron laser,'' Physical Review Letters 48, 475--477 (1982). DOI: 10.1103/PhysRevLett.48.475.

\bibitem{ref15} C. B. Schroeder, C. Pellegrini and P. Chen, ``Quantum effects in high-gain free-electron lasers,'' Physical Review E 64, 056502 (2001). DOI: 10.1103/PhysRevE.64.056502.

\bibitem{ref16} R. Bonifacio, N. Piovella, G. R. M. Robb and A. Schiavi, ``Quantum regime of free electron lasers starting from noise,'' Physical Review Special Topics--Accelerators and Beams 9, 090701 (2006). DOI: 10.1103/PhysRevSTAB.9.090701.

\bibitem{ref17} P. Kling, E. Giese, R. Endrich, P. Preiss, R. Sauerbrey and W. P. Schleich, ``What defines the quantum regime of the free-electron laser?,'' New Journal of Physics 17, 123019 (2015). DOI: 10.1088/1367-2630/17/12/123019.

\bibitem{ref18} P. Kling, E. Giese, C. M. Carmesin, R. Sauerbrey and W. P. Schleich, ``High-gain quantum free-electron laser: Emergence and exponential gain,'' Physical Review A 99, 053823 (2019). DOI: 10.1103/PhysRevA.99.053823.

\bibitem{ref19} P. Kling, E. Giese, C. M. Carmesin, R. Sauerbrey and W. P. Schleich, ``High-gain quantum free-electron laser: Long-time dynamics and requirements,'' Physical Review Research 3, 033232 (2021). DOI: 10.1103/PhysRevResearch.3.033232.

\bibitem{ref20} A. Bambini and A. Renieri, ``The free electron laser: A single-particle classical model,'' Lettere al Nuovo Cimento 21, 399--404 (1978).

\bibitem{ref21} A. Bambini, A. Renieri and S. Stenholm, ``Classical theory of the free-electron laser in a moving frame,'' Physical Review A 19, 2013--2025 (1979).

\bibitem{ref22} G. Dattoli, J. Gallardo, A. Renieri and M. Richetta, ``Quantum statistical properties of an FEL amplifier,'' IEEE Journal of Quantum Electronics 21, 1069--1072 (1985).

\bibitem{ref23} G. Dattoli, J. Gallardo and A. Torre, ``Quantum multimode analysis of the free-electron laser,'' Physical Review A 31, 3755--3760 (1985).

\bibitem{ref24} P. Kling and E. Giese, ``Quantum free-electron laser oscillator,'' Scientific Reports 16, 10521 (2026). DOI: 10.1038/s41598-026-45068-1.

\bibitem{ref25} F. Ciocci, G. Dattoli, A. Renieri and A. Torre, ``About the mathematical aspects of the FEL and their relationship to quantum optics,'' Physics Reports 141, 1--50 (1986).

\bibitem{ref26} M. S. Zolotorev and K. T. McDonald, ``Classical Radiation Processes in the Weizsäcker--Williams Approximation,'' arXiv:physics/0003096 (2000). \url{https://arxiv.org/abs/physics/0003096}.

\bibitem{ref27} G. Dattoli and A. Renieri, "Experimental and Theoretical Aspects of the Free-Electron Laser," in Laser Handbook, Vol. IV, edited by M. L. Stitch and M. Bass (Elsevier, 1985), pp. 1-140, especially Section 7, pp. 54-71. ISBN-13: 978-0444869272.

\bibitem{ref28} A. Proca, ``Sur la théorie ondulatoire des électrons positifs et négatifs,'' Journal de Physique et le Radium 7, 347--353 (1936). DOI: 10.1051/jphysrad:0193600708034700.

\bibitem{ref29} A. S. Goldhaber and M. M. Nieto, ``Photon and graviton mass limits,'' Reviews of Modern Physics 82, 939--979 (2010). DOI: 10.1103/RevModPhys.82.939.

\bibitem{ref30} G. Dattoli, ``A Bloch Like Model of the Free Electron Laser,'' Lettere al Nuovo Cimento 27, 247--251 (1980).

\bibitem{ref31} G. Baym, Lectures on Quantum Mechanics, W. A. Benjamin, New York (1969).

\bibitem{ref32} F. Bloch and A. Nordsieck, ``Note on the radiation field of the electron,'' Physical Review 52, 54--59 (1937). DOI: 10.1103/PhysRev.52.54.

\bibitem{ref33} G. Dattoli and A. Torre, ``A quantum mechanical point of view to perturbative problems in classical mechanics,'' Journal of Mathematical Physics 34, 5062--5071 (1993). DOI: 10.1063/1.530342.

\bibitem{ref34} J.-M.-C. Duhamel, ``Sur la méthode générale relative au mouvement de la chaleur dans les corps solides plongés dans des milieux dont la température varie avec le temps,'' Journal de l'École Polytechnique 14, cahier 22, 20--77 (1833).

\bibitem{ref35} F. J. Dyson, ``The radiation theories of Tomonaga, Schwinger, and Feynman,'' Physical Review 75, 486--502 (1949). DOI: 10.1103/PhysRev.75.486.

\bibitem{ref36} G. Dattoli, J. C. Gallardo and A. Torre, ``Time-ordering techniques and solution of differential difference equation appearing in quantum optics,'' Journal of Mathematical Physics 27, 772--780 (1986). DOI: 10.1063/1.527182.

\bibitem{ref37} G. Dattoli, H. Fares and S. Licciardi, ``Space charge and quantum corrections in free electron laser evolution,'' Nuclear Instruments and Methods in Physics Research A 976, 164176 (2020). DOI: 10.1016/j.nima.2020.164176, especially Eqs. (2), (5), (26), (36)--(37).

\bibitem{ref38} C. V. Raman and N. S. Nagendra Nath, ``The diffraction of light by high frequency sound waves: Part I,'' Proceedings of the Indian Academy of Sciences A 2, 406--412 (1935).

\bibitem{ref39} G. Dattoli, J. C. Gallardo, A. Renieri and A. Torre, ``Comment on `Photon antibunching in a free-electron laser','' Physical Review A 32, 3134 (1985). DOI: 10.1103/PhysRevA.32.3134.

\bibitem{ref40} E. Wigner, ``On the Quantum Correction for Thermodynamic Equilibrium,'' Physical Review 40, 749--759 (1932). DOI: 10.1103/PhysRev.40.749.

\bibitem{ref41} J. E. Moyal, ``Quantum mechanics as a statistical theory,'' Proceedings of the Cambridge Philosophical Society 45, 99--124 (1949). DOI: 10.1017/S0305004100000487.

\bibitem{ref42} W. P. Schleich, Quantum Optics in Phase Space, Wiley-VCH, Berlin (2001).

\bibitem{ref43} C. Moritz Carmesin, P. Kling, E. Giese, R. Sauerbrey and W. P. Schleich, ``Quantum and classical phase-space dynamics of a free-electron laser,'' Physical Review Research 2, 023027 (2020). DOI: 10.1103/PhysRevResearch.2.023027.

\bibitem{ref44} N. Piovella, M. M. Cola, L. Volpe, A. Schiavi and R. Bonifacio, ``Three-dimensional Wigner-function description of the quantum free-electron laser,'' Physical Review Letters 100, 044801 (2008). DOI: 10.1103/PhysRevLett.100.044801.

\bibitem{ref45} D. Meschede, H. Walther and G. Müller, ``One-atom maser,'' Physical Review Letters 54, 551 (1985). DOI: 10.1103/PhysRevLett.54.551.

\bibitem{ref46} G. Rempe, H. Walther and N. Klein, ``Observation of quantum collapse and revival in a one-atom maser,'' Physical Review Letters 58, 353--356 (1987). DOI: 10.1103/PhysRevLett.58.353.

\bibitem{ref47} H. Walther, B. T. H. Varcoe, B.-G. Englert and T. Becker, ``Cavity quantum electrodynamics,'' Reports on Progress in Physics 69, 1325--1382 (2006). DOI: 10.1088/0034-4885/69/5/R02.

\bibitem{ref48} E. T. Jaynes and F. W. Cummings, ``Comparison of quantum and semiclassical radiation theories with application to the beam maser,'' Proceedings of the IEEE 51, 89--109 (1963). DOI: 10.1109/PROC.1963.1664.

\bibitem{ref49} G. Dattoli, T. Hermsen, L. Mezi, A. Torre and J. C. Gallardo, ``A review of the theory of pulse propagation in the long bunch low gain regime,'' Nuclear Instruments and Methods in Physics Research A 272, 351--363 (1988), especially Eqs. (10) and (19).

\bibitem{ref50} G. Dattoli, P. L. Ottaviani and S. Pagnutti, ``High gain oscillators: Pulse propagation and saturation,'' Journal of Applied Physics 101, 103109 (2007).

\bibitem{ref51} P. Kling and E. Giese, ``Quantum free-electron laser oscillator: Supplementary Information,'' supplementary material to Scientific Reports 16, 10521 (2026), Section S1.1, Eqs. (S.1)--(S.13). Available with doi:10.1038/s41598-026-45068-1.

\bibitem{ref52} G. Dattoli, L. Giannessi and P. L. Ottaviani, ``Oscillator-amplifier free electron laser devices with stable output power,'' Journal of Applied Physics 95, 3211--3216 (2004). DOI: 10.1063/1.1645649, especially Eq. (7).

\bibitem{ref53} P. L. Kapitza and P. A. M. Dirac, ``The reflection of electrons from standing light waves,'' Proceedings of the Cambridge Philosophical Society 29, 297--300 (1933). DOI: 10.1017/S0305004100011105.

\bibitem{ref54} D. L. Freimund, K. Aflatooni and H. Batelaan, ``Observation of the Kapitza--Dirac effect,'' Nature 413, 142--143 (2001). DOI: 10.1038/35093065.

\bibitem{ref55} H. Batelaan, ``Colloquium: Illuminating the Kapitza--Dirac effect with electron matter optics,'' Reviews of Modern Physics 79, 929--941 (2007). DOI: 10.1103/RevModPhys.79.929.

\bibitem{ref56} S. Nehemia, R. Dahan, M. Shentcis, O. Reinhardt, Y. Adiv, K. Wang, O. Be'er, Y. Kurman, X. Shi, M. H. Lynch and I. Kaminer, ``Observation of the Stimulated Quantum Cherenkov Effect,'' arXiv:1909.00757 (2019).

\bibitem{ref57} J. Walsh, B. Johnson, G. Dattoli and A. Renieri, ``Undulator and \v{C}erenkov Free-Electron Lasers: A Preliminary Comparison,'' Physical Review Letters 53, 779 (1984). DOI: 10.1103/PhysRevLett.53.779.

\bibitem{ref58} S. Tomin, E. Schneidmiller and W. Decking, ``First measurement of energy diffusion in an electron beam due to quantum fluctuations in the undulator radiation,'' Scientific Reports 13, 1605 (2023). DOI: 10.1038/s41598-023-28813-8.

\bibitem{ref59} G. Geloni, V. Kocharyan and E. Saldin, ``On Quantum Effects in Spontaneous Emission by a Relativistic Electron Beam in an Undulator,'' Proceedings of FEL2012, paper MOOC03 (2012). \url{https://proceedings.jacow.org/FEL2012/papers/mooc03.pdf}.

\bibitem{ref60} G. Dattoli, P. L. Ottaviani and S. Pagnutti, Booklet for FEL Design: A Collection of Practical Formulae, published by ENEA - Edizioni Scientifiche, Section 3.2, Eqs. (3.10)-(3.12). \url{http://fel.enea.it/booklet/pdf/Booklet_for_FEL_design.pdf}

\end{thebibliography}
\end{document}